# CHAPTER 5

# SMALL QUANTUM DOTS OF DILUTED MAGNETIC III-V SEMICONDACTOR COMPOUNDS

Liudmila A. Pozhar

*PermaNature, Birmingham, AL 35242*

Home Address: 149 Essex Drive, Sterrett, AL 35147

Tel: (205) 678-0934

E-mail: lpozhar@yahoo.com; pozharla@yahoo.com

Running Title: **Small QDs of III-V DMS**

## SUMMARY

In this chapter quantum many body theoretical methods have been used to study properties of GaAs - and InAs - based, small semiconductor compound quantum dots (QDs) containing manganese or vanadium atoms. Interest to such systems has grown since experimental synthesis of nanoscale magnetic semiconductors, that is, nanoscale semiconductor compounds with enhanced magnetic properties. This enhancement is achieved by several methods, and in particular by doping common semiconductor compounds with "magnetic" atoms, such as Mn or V. Experimental studies indicate that the electron spin density in the case of thin nanoscale "magnetic" semiconductor films and QDs may be delocalized. As described in this chapter, quantum many body theory-based, computational synthesis (*i.e.*, virtual synthesis) of tetrahedral symmetry GaAs and InAs small pyramidal QDs doped with sabstitutional Mn or V atoms proves that such QDs are small "magnetic" molecules that indeed, possess delocalized and polarized electron spin density. Such delocalization provides a physical mechanism responsible for stabilization of these nanoscale molecular magnets, and leads to the development of what can be described as spin-polarized "holes" of the electron charge deficit. In some QDs, numerical values of the electron spin density distribution are relatively large, indicating that such semiconductor systems may be used as nanomaterials for spintronic and magneto-optical sensor applications.

# 1. INTRODUCTION

Since a discovery of ferromagnetism in bulk Mn-doped GaAs and InAs systems (1- 4), III-V - and II-VI - based diluted magnetic semiconductor (DMS) have been vigorously studied both by experimental and theoretical means. Such systems exhibit a variety of electronic and magneto-optical properties defined by carrier-mediated ferromagnetism, including spin polarization of holes, turning on and off the magnetic phase by changing hole concentration at constant temperature, hysteresis of Hall resistance as a function of an external magnetic field and gate voltage below the Curie temperature (Tc), electric-field assisted magnetization reversal, magnetization enhancement by circularly polarized light, current-induced magnetization switching, a strong increase of resistance with lowering temperature, colossal negative magnetoresistance, current-induced domain wall motion, *etc*. (see Refs. 5 - 7 and references therein). These properties are exceptionally important in materials for quantum electronics, spintronics and quantum information processing, as they allow control of carriers' spin generation, relaxation and deflection, and magneto-optical properties of the materials, using not only external magnetic fields, but also external electric fields and/or current, light, carrier concentration, and quantum confinement effects.

Unfortunately, bulk DMS systems possess very low Curie temperature ($T_C$) that heavily limits their technological use. Thus, synthesis of DMS-based structures with room temperature $T_C$ and higher has become one of the central issues of application of DMS-based materials in novel spintronics and quantum information processing technologies. In particular, following the existing practical experience and the p-d Zener model-based predictions (8, 9), state-of-the-art low temperature molecular beam epitaxy (MBE) and other thin film synthesis methods, layered device development techniques, and numerous methods of synthesis of quantum wire (QW) - and quantum dot (QD) - based nanostructures have been established over the years (see, for example, Ref. 10 - 16) to realize low dimensional DMS-based nanostructures where DMS components possess enhanced $T_C$ up to 173 K (17 - 26). Recently, room temperature ferromagnetism was reported in Mn-doped InGaAs and GaAsSb nanowires (27), and MBE-based formation of (Ga,Mn)As crystalline nanowires with Tc=190 K was achieved (28). However, reached values of $T_C$ for ingenious layered structures based on Mn-doped (Ga,Mn)As, that is expected to be the highest $T_C$ DMS among III-V candidates, are significantly lower than those predicted by the existing theoretical models of hole-mediated ferromagnetism in bulk, spatially homogeneous Mn-doped GaAs and InAs systems. In particular, the *p-d* Zener model (8, 9) predicts $T_C$=345 K for a spatially homogeneous sample $Ga_{0.8}Mn_{0.2}As$ with high Mn concentration from Ref. 11, while the observed $T_C$ for that structure is 118 K. Because the p-d Zener model is designed to describe DMSs in the low impurity concentration limit, it also fails to predict (even qualitatively) an increase in $T_C$ of samples with high Mn concentration (11). Similarly, an improved approach (29) that uses a semi-phenomenological model predicts unrealistic values of $T_C$ above 500 K in structures with remote doping, such as heterostructures of Ref. 24. [In such heterostructures impurities are introduced as a so-called δ layer on one side of a heterojunction, so impurities and carriers are not in the same channel. This provides for the carrier mobility up to three orders of magnitude higher than that in heterostructures where impurities and carriers share the channel.]

It is important to note that the exiting theoretical approaches have been developed to describe ferromagnetism (FM) in bulk, homogeneous DMDs with low impurity concentration (30, 31). Correspondingly, they incorporate assumptions that are not applicable to quantum-confined systems and nanoscale heterostructures, such as those composed of layered films, QWs and QDs of a few nanometers in linear dimensions. In such systems carrier motion is quantized in the direction(s) of confinement, which is not incorporated into the models' formalism. Moreover, a small number of atoms (2 to 4, in the case of a confinement with at least one linear dimension of about 1 nm) require consideration of many-particle quantum problem. In the absence of a rigorous, first principle quantum theory of solids, and in particular semiconductors, necessary to fully account for quantum effects, the semi-phenomenological band theory of semiconductors was modified to produce numerous models designed to quantify observed electronic structure and properties of semiconductor films, quantum wells, thick QWs, large QDs, and their heterostructures (30, 32, 33).

One of such models due to Zener (34), called the *p-d* Zener model, has been very successful in the case of bulk DMSs (8, 9), and with some modifications has also been successfully applied to thick DMS films and layered systems (see, for example, Refs. 31 and references therein). This model makes use of the fact that in DMSs the Fermi energy $E_F$ lies, as a rule, within the majority *t*-band of the "magnetic" impurity (such as Mn atoms) which is $t_{2g}$ symmetric. Moreover, *d*-states of magnetic impurity atoms lie below the valence *p*-states of the host semiconductor structure, like in (Ga,Mn)As. Such *p*- and *d*-states hybridize pushing the majority valence states to high energies, and the minority valence states to lower energies. This mechanism, called kinetic *p-d* exchange, leads to appearance of holes in the valence band, and thus to generation of a significant magnetic moment μ per an impurity atom (μ = 5$μ_B$ in the case of Mn impurity, where $μ_B$ is the Bohr magneton) leading to hole-mediated ferromagnetism. At the same time, the majority valence band becomes spin-polarized with a much smaller magnetic moment per an impurity atom (equal to -$μ_B$ per a Mn atom) due to weak *d-d* interactions between impurity electrons. Thus, the majority (*d*) - majority (*p*) orbit hybridization (the *p-d* exchange) produces FM coupling that leads to a decrease in the total energy of the system.

Importantly, the *p-d* Zener model exploits the virtual crystal and molecular field approximations, the limit of low hole and impurity concentrations (with the hole concentration much smaller than that of the impurities), and an assumption of weak *p-d* coupling (35). In particular, when the *p-d* hybridization is large, the coupling becomes strong, and both approximations become invalid (36, 37). Also, the *p-d* Zener model is expected to fail in the case of nanometer-thin DMS films, thin QWs and small QDs where the concentration of impurity atoms may reach tens of percent. Moreover, for DMS systems possessing large characteristic dimensions, but heterogeneous in nature (that is, featuring regions of the higher and smaller impurity concentration, or different magnetic phases) this model is also invalid. Indeed, the *p-d* Zener model predicts the Curie temperature $T_C \approx 345°$ K in the case of a FM-homogeneous $Ga_{1-x}Mn_xAs$ film with the Mn concentration *x* over 10%, while the experimental value is approximately 118° K (12). According to this model, $T_C$ increases with the increasing hole concentration. However, when the hole concentration exceeds that of the impurities, experimentally observed $T_C$ of 5 nm thin $Ga_{1-x}Mn_xAs$ from Ref. 11 tends to decrease with an increase in the hole concentration. Moreover, the *p-d* Zener model does not account for many other experimental data (14, 38 - 41) on magnetic properties of DMS films. There were attempts to modify this model to account for new experimental observations (18, 42), but they could not accommodate for emerging experimental evidence indicating that FM mechanisms other than the *p-d* Zener exchange, significantly contributed to magnetic properties of DMSs.

Thus, other semi-phenomenological models, such as an impurity band (IB) model have been developed (43). Yet, even in the case of relatively thick DMS films of Ref. 14, the major issue concerning the hole-mediated FM mechanism remains unresolved. In particular, it is not quite clear where the mediating carriers reside – in a disordered valance band (VB), as stated in Refs. 6 and 8, or in the impurity band that may be detachable from the host VB and retaining *d*-orbit properties of the impurities, such as Mn (43 - 50).

More sophisticated theoretical approaches steaming from the first-principle theoretical basis have been suggested to explain non-conventional origin of ferromagnetism in and complex magnetic properties of DMSs. Most likely, the earliest of them was an approach that used scattering theoretical methods to solve Schrödinger equation (51 - 53) for electrons in a periodic system. However, another approach due to Korringa, Kohn and Rostoker (KKR) introduced in Refs. 54 - 56 became more popular due to its simplicity and mathematical transparency. This approach uses the density functional theory (DFT) mathematical foundation (54) and a smart choice of a reference system to simplify a general DFT procedure of band structure calculations. Numerous KKR-based models were developed and applied to periodic systems with localized perturbations (57 – 65). The KKR method itself was further improved when the problem of determining the charge density $n(\mathbf{r})$ was solved (66, 67) in terms of the single particle Green's function $G(\mathbf{r}_1, \mathbf{r}_2)$ introduced as the solution of the inhomogeneous Kohn-Sham equation of the DFT with the δ-function source, and the algebraic Dyson equation used to relate the Green's function of structural defects to the Green's function of the ideal crystal (68). Further advance of the KKR method is related to inclusion of non-spherical interaction potentials using a shape function technique to model non-spherical parts of the potential, and using the Lippmann-Schwinger equation to develop an iterative scheme of determination of the Green's functions from the Dyson equation (68 – 72). A concept of a coherent potential (73, 74) was incorporated to make KKR-based methods applicable to disordered systems and giving rise to so-called the coherent potential approximation (CPA). The use of the local force theorem (75) enabled further successful applications of KKR-CPA models to disordered magnetic systems. According to this "theorem", in the case of a frozen ground state potential and small perturbations in the charge and magnetization densities of a system, a variation of the total energy of the system is equal to the sum of the single particle energies over all occupied energy states. This theorem simplified evaluation of the exchange interaction energy between two magnetic atoms (76).

Simplification of the exchange-correlation functional by the use of the local density approximation (LDA) made it practical to apply DFT-based methods to calculate the energy level structure of large molecules and the band structure of solids. However, LDA failed in the case of transition metal compounds and DMSs, where it predicted a partially filled $d$-band with metallic character of the electronic level structure (77). Thus, other adjustments of DFT-based methods were introduced, including the most popular self-interaction correction LDA (SIC-LDA) and LDA+U (78, 79) methods, with U being the Hubbard potential, to reach at least partial success in predicting the band structure and $T_C$ of strongly correlated systems, and in particular, DMSs. In GaAs systems with Mn impurities the Hubbard potential accounts for Coulomb correlation effects that have to be incorporated to calculate more accurately the band structure of DMSs (79), and also leads to an increase in hole delocalization and a decrease in the $p$–$d$ Zener interaction (78). It also inhibits the double exchange and pushes the $d$-states to lower energies (80). As a result, in LDA+U approximation the mean-field Curie temperature is a linear function of the carrier concentration (81). Despite important results, it is not clear whether further improvement in prediction of the band structure and magnetic properties of DMS is possible in the framework of LDA and its modifications.

Recently, a more general, spin-polarized version of a DFT-based generalized gradient approximation (σGGA) and its modification by inclusion of the Hubbard potential (σGGA+U) were used (82) to calculate geometry, and electronic and magnetic properties of Mn atoms incorporated in bulk GaAs. The electron-electron interactions were described using the Perdew–Burke–Ernzerhof exchange–correlation potential (83), and electron-ion interactions were described by ultrasoft pseudopotentials (84). It was proved that inclusion of the Hubbard potential into calculations (σGGA + U) leads to contraction of geometrical parameters by about 2% as compared to those calculated without the Hubbard potential by σGGA method. The σGGA + U calculations of Ref. 78 have used the value 4.0 eV for the Hubbard potential, which is usually used in LDA+U calculations, and which is chosen to be close to the value U=3.5 eV obtained on the basis of experimental data for photoemission for $Ga_{1-x}Mn_xAs$ with small values of the impurity concentrations $x$ (85). In both cases of σGGA and σGGA+U, a Mn impurity atom introduces a $d$-hole, and the majority spin state of the impurity lies at 0.25 eV above the Fermi level. These theoretical data correlate with experimental observations for bulk GaAs systems with low concentration of Mn, and constitute improvement of the previous LDA+U results.

It has become obvious, that being a non-variational theory by its nature (86), DFT cannot be applied successfully to predict properties of transition metal semiconductor (87) and magnetic systems without *ad hoc* schemes (such as KKR), uncontrolled approximations (such as LDA, GGA or σGGA), pseudopotentials, and adjustable parameters, such as the Hubbard potential U. All these improvements and adjustments are incorporated to match experimental observations, and are not derivable from the first principles. They are necessary because in the framework of DFT one cannot introduce a self-consistent scheme of controlled approximations to the "real" exchange-correlation functional. Even more such *ad hoc* manipulations are required to apply DFT-based methods to strongly correlated nanoscale systems, such as layered DMS systems, or their QW and/or QD assemblies. Moreover, one cannot predict results of any of DFT-based theoretical schemes, until *ad hoc* theoretical approximations for the exchange-correlation functionals and adjustable parameters steaming from experimental data are incorporated into such theoretical schemes.

During two recent decades, another class of theoretical approaches to the electronic structure-property correlations in spatially inhomogeneous and nanoscale magnetic systems has emerged. These approaches have been developed originally to describe quantum phenomena (88 – 94), such as Coulomb blockade, in large nanoscale systems, and are focused on electron dynamics. They use a rigorous quantum statistical mechanical basis (see, for example, Refs. 95 and 96), but include numerous uncontrolled approximations, *ad hoc* assumptions and intuitive models, such as those explored in Refs. (97, 98). These approaches has received significant attention in literature in conjunction with emerging quantum computing technologies (99), because spin states of electrons residing on magnetic QDs are considered as the most stable realization of quantums of information qubits (100 – 102). Being derived semi-phenomenologically from a theoretical basis developed either for bulk systems, or for simple quantum mechanical models, these approaches (103 – 111) tend to provide physically incorrect predictions even for mesoscopic cases, such as tunneling junctions, because they do not include adequate consideration of the electron spin interactions, quantum confinement effects, QD-to-QD and QD-to-environment coupling. However, such interactions, effects and coupling are responsible for the major mechanisms of ferromagnetism in DMSs. By their nature, such models do not allow first-principle predictions of the electronic structure and magneto-electronic properties of layered or nanoscale magnetic systems. Fortunately, in the case of small nanoscale systems one can exploit rigorous methods of quantum statistical mechanics (QSM) directly using QSM theory-based "quantum chemistry" software packages, such as GAMESS, GAUSSIAN or Molpro, as described in the following section.

## 2. VIRTUAL SYNTHESIS OF SMALL QUANTUM DOTS OF III – V MAGNETIC SEMICONDUCTOR COMPOUNDS

Latest resonant tunneling spectroscopy studies of a variety of GaMnAs surface layers (112) have proven that GaAs band structure is not significantly affected by the presence of Mn substitutional defects for a range of Mn concentrations from 6% to 15%. It appeared, that the GaAs valence band (VB) does not merge with the impurity band, and that the exchange splitting of VB is in the range of several millielectronvolts even for the layers with $T_C$ as high as 154º K. At the same time, the ferromagnetic state was more pronounced than that of bulk (Ga,Mn)As and (Ga,Mn)As (26). Moreover, new model studies (113) show that similar to atoms and nuclei QDs of DMS exhibit formation of electron "shells", and that their magnetism depends on shell occupancy. There exist nanoscale potential fluctuations in (Ga,Mn)As/GaAs heterostructures that lead to formation of electrostatic QDs (114). Transversal Kerr effect spectroscopy studies (115) identified that MnAs inclusions of 10 nm to 40 nm in linear dimensions were responsible for a strong resonant band in the energy range from 0.5 eV to 2.7 eV in GaMnAs and InMnAs layers formed by laser ablation on GaAs and InAs surfaces. Observed transitions from quasi-two dimensional (2D) to three dimensional (3D) $In_{0.85}Mn_{0.15}As$ structures develop gradually and are rather slow even at 270º C (116). At the same time, electron spin relaxation time in MnAs nanoparticles formed in GaAs lattice was found to be as long as 10 μs at 2º K (117), in contrast to known relaxation times of the order 100 ns in other QD structures and several picoseconds in bulk semiconductor systems (102).

These latest experimental data are in contradiction with the existing conventional models of mechanisms responsible for magnetism in and the band structure of nanoscale DMS systems, as described in Sec. 1. In particular, the major assumption that holes occupy GaAs-like valence band used to obtain successful predictions (118) of magnetization of GaMnAs as a function of the magnetic field at $T_C$ contradicts to experimental results of Ref. 112 and modeling results of Ref. 113. Thus, with advance of spintronics and quantum computing, more accurate QSM-based methods are required to evaluate spin entanglement and decoherence, and spin transport properties in small QDs and their heterostructures.

In this chapter the first principle QSM methods free from any assumptions concerning mechanisms and the band structure of the studied systems have been used to synthesize virtually a range of 14-atomic QDs of GaAs and InAs containing one or two Mn or V atoms and study their electronic and magnetic properties. The virtual synthesis method of Refs. 119 – 121 as realized by the GAMESS software package and discussed in Chapter 3 have been applied to minimize the total energy of several atomic clusters composed of tetrahedral symmetry elements (14 - atomic pyramids) of the GaAs and InAs zincblende lattices as described in Chapter 4 to obtain $Ga_{10}As_4$ and $In_{10}As_4$ molecules. Original pyramidal frames of these clusters are built of 10 Ga or In atoms, and four As atoms are placed at ¼ of the cube body diagonals in their bulk zincblende lattices (that is, inside of the pyramidal frames composed of the Ga or In atoms). The initial covalent radii of Ga, In and As atoms in these clusters have been adopted from experiment: 1.26 Å, 1.44 Å and 1.18 Å, respectively. [Properties of these molecules are discussed in Chapter 4 in details.] Such clusters were synthesized at conditions mimicking quantum confinement (when spatial constraints were applied to the centers of mass of clusters' atoms) and in "vacuum", that is, in the absence of any constraints (external fields or "foreign" atoms interacting with the clusters' atoms). At the next step, one or two As atoms have been replaced by one or two vanadium or manganese atoms, without any changes to the positions of the remaining Ga or In and As atoms. The total energy of the so built "diluted magnetic semiconductor" clusters has been minimized in the presence and in the absence of spatial constraints applied to the clusters' atoms to obtain new pre-designed and vacuum molecules, respectively. Note again, that spatial constraints applied to the centers of mass of the clusters' atoms have been incorporated to reflect effects of quantum confinement on the molecules synthesized in such a confinement. The total energy of the clusters so built have been minimized using Hartree-Fock (HF) and restricted open shell Hartree Fock (ROHF), leading to the emergence of pre-designed molecules $Ga_{10}As_3V$, $Ga_{10}As_2V_2$, $In_{10}As_3V$, $In_{10}As_2V_2$, and $In_{10}As_3Mn$. Thus, the Schrödinger equation has been solved numerically using GAMESS software to obtain the ground state of the corresponding molecules in the presence of the boundary conditions realized as the spatial constraints applied to positions of the centers of mass of the clusters' atoms. Once the pre-designed molecules were obtained, the corresponding "vacuum" molecules have been developed by relaxing the spatial constraints applied to the atomic positions in the pre-designed clusters, and subsequent optimization (solving the corresponding Schrödinger equations in the absence of the spatial constraints). Electronic and magnetic properties of the virtually synthesized molecules are discussed below in this chapter. Ground state details of the constitutive atoms are listed in Table 1.

TABLE I. Ground States of Semiconductor Compound Atoms

| Atom | Electronic Configuration | Atomic Term | Nuclear Spin | Nuclear Magnetic Moment, in $\mu_P$ Units |
|---|---|---|---|---|
| $^{31}$Ga | [Ar]$3d^{10}4s^24p$ | $^2P_{1/2}$ | 3/2 | 2.0166 |
| $^{49}$In | [Kr] $4d^{10}5s^25p$ | $^2P_{1/2}$ | 9/2 | 5.5340 |
| $^{33}$As | [Ar] $3d^{10}4s^24p^3$ | $^4S_{3/2}$ | 3/2 | 1.4395 |
| $^{25}$Mn | [Ar]$3d^54s^2$ | $^6S_{5/2}$ | 5/2 | 3.4687 |
| $^{23}$V | [Ar] $3d^34s^2$ | $^4F_{3/2}$ | 7/2 | 5.1574 |

## 3. PRE-DESIGNED AND VACUUM MOLECULES $In_{10}As_3Mn$

The case of the $In_{10}As_3Mn$ molecules is very special from several points of view. In particular, findings obtained in the process of virtual synthesis of such molecules may help shed light of the origin of some controversies between the existing semi-phenomenological theoretical predictions and experimental observations specific to InAs nanoscale systems.

In contrast to the majority of the studied zincblende-derived molecules (see Chapter 4 and subsequent sections of the current chapter for details), the shape of the vacuum $In_{10}As_3Mn$ molecule visibly deviates from pyramidal shape of its pre-designed counterpart (Figs. 1 and 2).

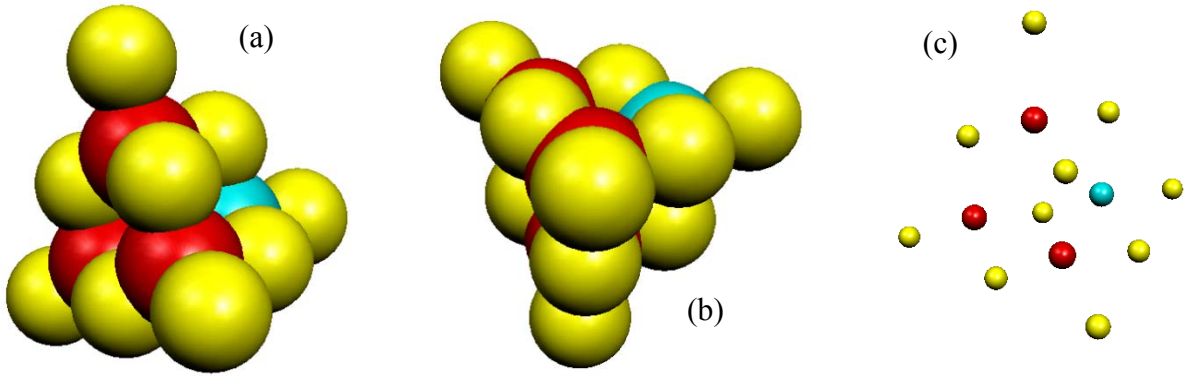

Fig. 1.     (Color online) The pre-designed pyramidal molecule $In_{10}As_3Mn$: (a) front view; (b) top view; (c) atomic positions. Indium atoms are yellow, As red, and Mn blue. In (a) and (b) all dimensions are approximately to scale; atomic dimensions roughly correspond to their respective covalent radii. In (c) atomic dimensions are reduced.

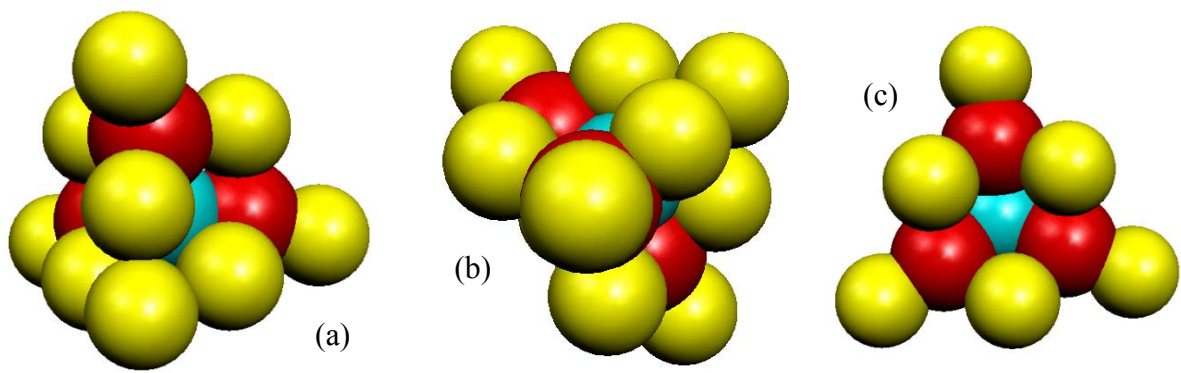

Fig. 2. (Color online) The vacuum molecule $In_{10}As_3Mn$. (a) and (b): side views; (c) front view. Indium atoms are yellow, As red, and Mn blue. In (a) and (b) all dimensions are approximately to scale; atomic dimensions roughly correspond to their respective covalent radii.

As compared to the pre-designed pyramid of Fig. 1, in the vacuum molecule In atoms moved somewhat closer toward each other and Mn atom (Fig 2a). This is enough for the vacuum molecule to lose the pyramidal shape. The position adjustment allows stabilization of the vacuum $In_{10}As_3Mn$ molecule in the absence of any spatial constraints and external fields. The electron charge density distribution (CDD) and molecular electronic potential (MEP) of these molecules are shown in Figs. 3 and 4.

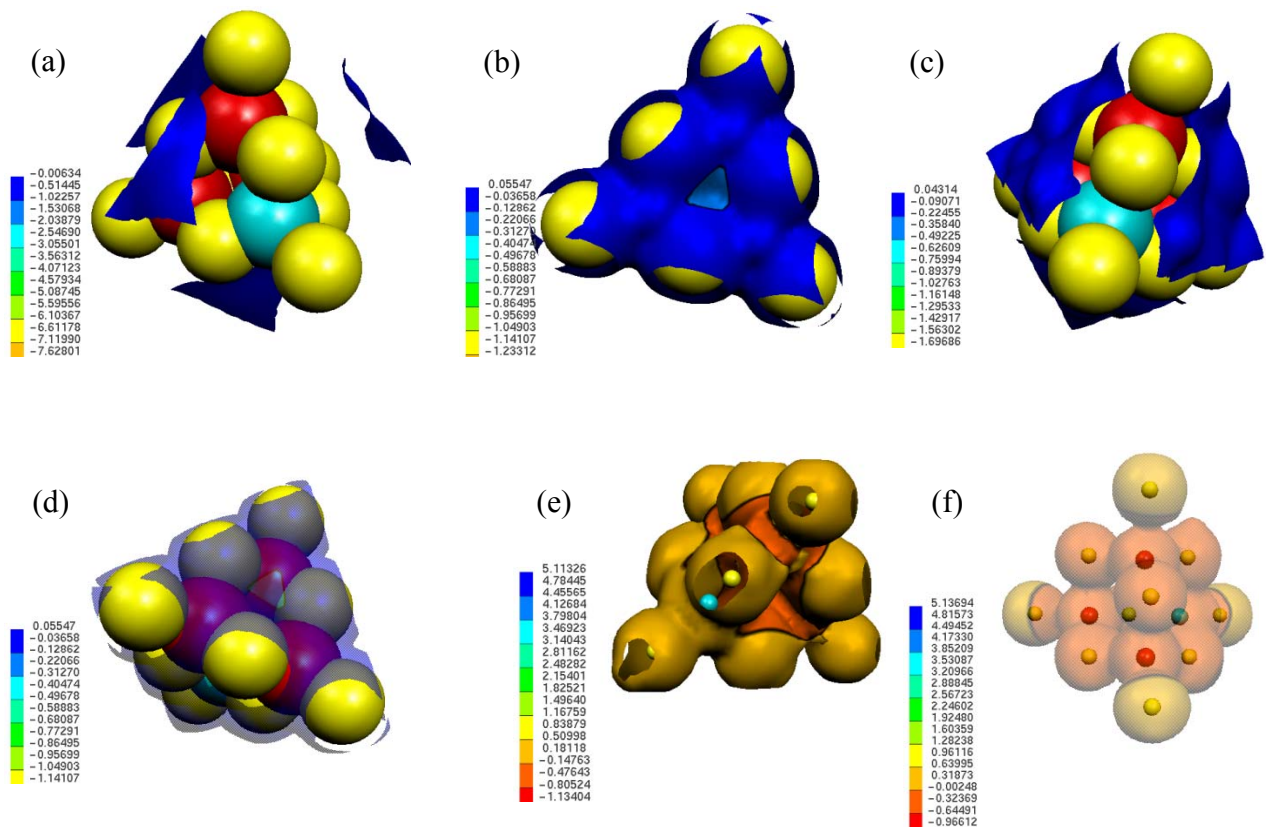

Fig. 3. The molecular electrostatic potential (MEP) of the pre-designed molecule $In_{10}As_3Mn$ for several isosurfaces of the electron charge density distribution (CDD) calculated for fractions (isovalues) of the CDD maximum value 4.25145 (arbitrary units): (a) 0.001; (b) 0.02; (c) and (d) 0.05; (e) 0.08; (f) 0.1. The color coding scheme is shown in each figure. In atoms are yellow, As red and Mn blue. All dimensions in (a) to (d) are to scale; atomic dimensions roughly correspond to the atoms' covalent radii. In (e) and (f) atomic dimensions are reduced to show the MEP surface structure.

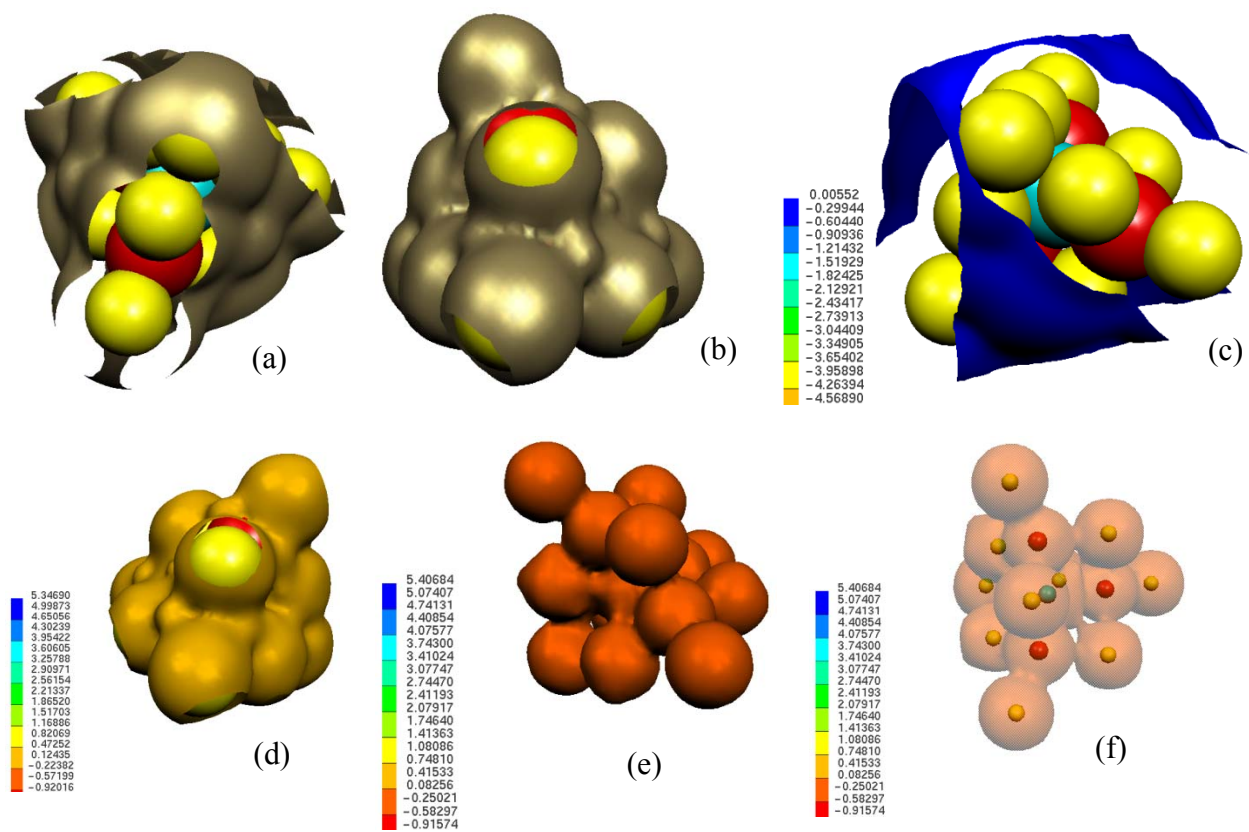

Fig. 4.  The electron charge density distribution [(a) and (b); CDD], and molecular electrostatic potential (MEP) of the vacuum molecule In$_{10}$As$_3$Mn for several isosurfaces of the CDD calculated for fractions (isovalues) of the CDD maximum value 4.61805 (arbitrary units). (a) and (b): 0.01 and 0.05, respectively; (c) to (f): MEP for CDD isovalues 0.001, 0.05, 0.1 and 0.1, respectively. The color coding scheme for MEP surfaces is shown in each figure. Indium atoms are yellow, As red and Mn blue. All dimensions in (a) to (d) are to scale; atomic dimensions roughly correspond to the atoms' covalent radii. In (e) and (f) atomic dimensions are reduced to show the MEP surface structure.

Both In$_{10}$As$_3$Mn molecules have similarly shaped CDD and MEP surfaces, and almost the same maximum and minimum values of CDD. However, the CDD and MEP surfaces of the vacuum molecule reflect significant loss of tetrahedral symmetry (compare the surfaces in Fig. 3a and Fig. 4c, for example). In both cases electron charge is delocalized, spread through the space occupied by the molecule and reaches beyond it. The electron charge density still is about 0.001 of the CDD maximum values at the distances about 2 covalent radii of In atoms from the molecular "surfaces" (Fig. 4a). Both molecules exhibit electron charge accumulation near their "surface" regions and nearby space on both "sides" of their surfaces. In the case of the pre-designed molecule MEP values at the distances from about 2 covalent radii of In atoms to the "surface" of the molecule as defined by its atomic covalent radii are more negative (Fig. 3a) than those closer to the "surface" of the molecule on its both "sides" (Figs. 3b to 3d), and the electron charge is relatively smoothly distributed in this region. Only well "inside" of the molecule MEP values become more negative than those in the space outside the molecule. Thus, despite of its total charge equal to zero, the pre-designed molecule may be characterized experimentally as a "shell" of delocalized negative charge surrounding the molecule at separations of about 2 covalent radii of In atoms from the "surface" of the molecule, and delocalized positive charge spread near the "surface on its both sides where there are regions of lesser values or zero electron charge. In the case of the vacuum molecule the electron charge is closer to the molecular "surface" on its both "sides", so the MEP values are more negative in the inner molecular regions, signifying that more electron charge is kept closer to the molecular "surface" and more of it is kept "inside" the molecule.

In the case of the pre-designed molecule the dipole moment is almost twice as large as that of the vacuum molecule (Table 2). Interestingly, the vacuum molecule is a ROHF septet whose uncompensated spin magnetic moment is equal to $7\mu_B$ (where $\mu_B$ denotes Bohr's magneton), and uncompensated electron spins delocalized in molecular orbits derived from $3d$ orbits of In atoms (Fig. 5). At the same time, the pre-designed molecule is a HF singlet with all electron spins compensated and the total spin magnetic moment equal to zero. The spin density distribution (SDD) of the vacuum molecule reaches into the space far outside of the molecule, in

TABLE II. Ground State Data for the Studied RHF/ROHF InAs- and GaAs-Based Molecules with One and Two Vanadium Atoms

| Molecule | Spin Multiplicity | RHF/ROHF Ground State Energy (Hartree) | RHF/ROHF Direct Optical Transition Energy (eV) | Dipole Moment (Debye) |
|---|---|---|---|---|
| $In_{10}As_3Mn$ | 1 | -2003.6919187179 | 3.9076 | 7.93980 |
| $In_{10}As_3Mn$* | 7 | -2004.2390957056 | 1.2436 | 3.46941 |
| $In_{10}As_3V$ | 9 | -1971.4803877865 | 0.0571 | 4.12400 |
| $In_{10}As_3V$* | 5 | -1971.4252752330 | 1.2653 | 3.26400 |
| $Ga_{10}As_3V$ | 3 | -2660.4273232664 | 1.2626 | 1.38721 |
| $Ga_{10}As_3V$* | 3 | -2660.5087787590 | 1.0585 | 4.56927 |
| $In_{10}As_2V_2$ | 11 | -2035.9308103524 | 0.1551 | 3.44417 |
| $In_{10}As_2V_2$* | 1 | -2036.0936312586 | 3.6082 | 3.27030 |
| $Ga_{10}As_2V_2$ | 7 | -2724,8716270248 | 0.8735 | 3.58697 |
| $Ga_{10}As_2V_2$* | 9 | -2724.9370304183 | 1.5837 | 2.37091 |

The star * denotes vacuum molecules.

agreement with the CDD and MEP data (Fig. 4). Thus, the vacuum molecule $In_{10}As_3Mn$ is a nanoscale ferromagnet with the rather large spin magnetic moment brought about by delocalization of electrons of the 3$d$ orbits of In atoms due to $d$-electrons of the Mn atom. One should note here, that this effect is not only due to Mn atom included in the molecule instead of an As atom. Rather, it is an integrated quantum Coulomb effect that is typical for nanoscale objects composed from a few atoms possessing 3$d$ electrons at conditions where there are no other electrons or atoms to help keep electrons in their octet rule-derived molecular orbits. To stabilize such systems, the octet rule must be violated, and electron charge delocalized throughout the system and even in the space surrounding the system. During this delocalization the electron spin components may or may not be fully compensated giving rise to molecules with the non-zero, and sometimes large, total magnetic moment.

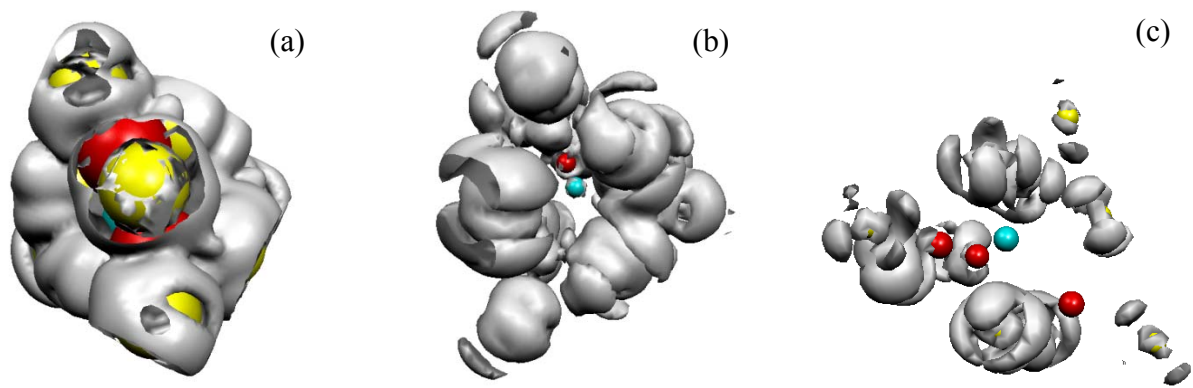

Fig. 5. Isosurfaces of the spin density distribution (SDD) of the vacuum molecule $In_{10}As_3Mn$ corresponding to the fractions (a) 0.0002, (b) 0.001 and (c) 0.005 of the SDD maximum value (not shown). Indium atoms are yellow, As red and Mn blue. In (a) all dimensions are to scale, with atomic sizes roughly corresponding to the atoms' covalent radii. In (b) and (c) atomic dimensions are reduced to show the SDD surface structure.

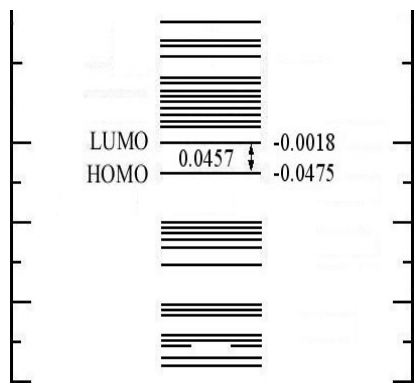

Fig. 6. The electronic energy level structure (ELS) in the HOMO-LUMO region of the vacuum In10As3Mn molecule. The HOMO and LUMO energy values, and the optical transition energy (OTE), are in Hartree (H) units.

The electronic energy level structure (ELS) of both molecules reflects the loss of tetrahedral symmetry due to the presence of the substitution Mn atom (Fig. 6) with its electronic configuration different from that of As ones. The vacuum molecule features bunches of closely lying orbits both in the region of the highest occupied and that of the lowest unoccupied orbits (HOMO and LUMO, respectively). Nevertheless, the ground states of both molecules correspond to deep minimum of the total energy, and their ELSs exhibit large OTEs (Table 2). The octet rule is violated, as about 240 electrons - many more than the number of usual "valence" ones - contribute to formation of the molecular orbitals (MOs) of these molecules. The nature of the MOs confirms the violation of the octet rule, and some of the MOs are similar to those introduced and studied in Ref. 122 for small non-stoichiometric atomic clusters. In the case of the pre-designed molecule, the bonding HOMO 118, which is the third occupied MO counting from the uppermost HOMO 120, is a *pd*-type MO derived from several major contributions: (1) 5*p* atomic orbits (AOs) of In atoms hybridized with 3*d* AOs of Mn, (2) 4*p* AOs of As atoms hybridized with 5*p* AOs of In atoms and 3*d* AOs of Mn atom, and (3) large *pp*- and small *pd*-ligand bonding of In atoms between themselves (Figs. 7a and 7b). These AOs also contribute to the rest of MOs in the near HOMO – LUMO region, including HOMO 119 and the proper HOMO 120 (Figs. 7c to 7e). In the latter case, contributions from 4*d* AOs of In atoms are significant. Ligand (In - In) bonding, primarily of *pp*-type and similar to that of π-type studied in Ref. 122, can be observed in all cases. The ligand bonding is mediated by As and Mn atoms. All three HOMOs in Figs. 7a to 7f are bonding MOs, which is a further illustration of stability of the pre-designed singlet. The LUMO 122 is also a bonding MO similar in nature to those of the HOMOs, but contributions from 4*d* AOs of In atoms here are negligibly small. In all cased In atoms bond to Mn via hybridization of their 5*p* AOs and 3*d* AOs of Mn, in agreement with Zener's assumption (34). However, in the pre-designed $In_{10}As_3Mn$ molecules, Mn and As atoms do not bond directly: their bonding is always mediated by 5*p* AOs of In atoms, in contrast to Mn-As 4*p*- bonding suggestion of Ref. 79.

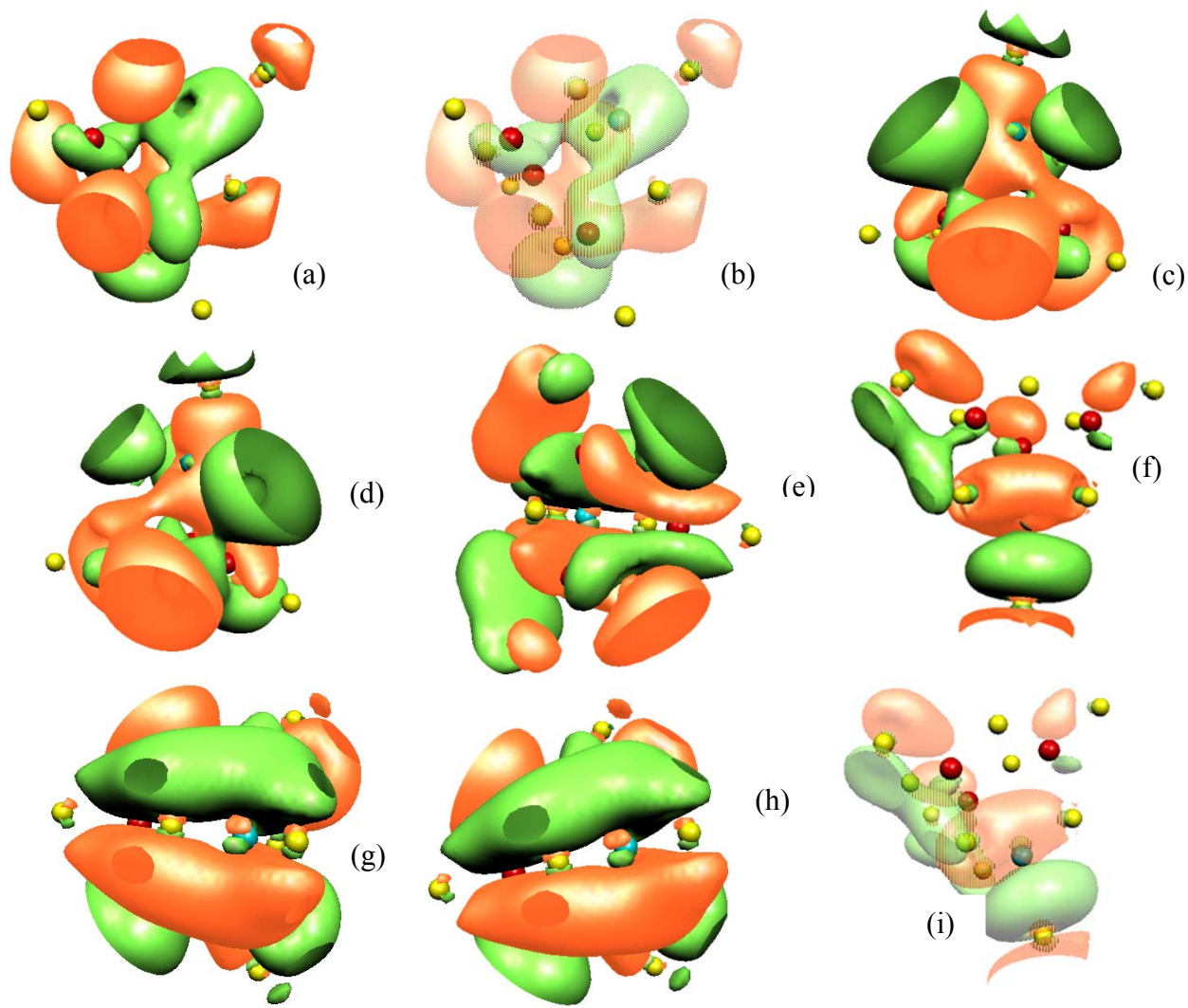

Fig. 7. (Color online) Pre-designed $In_{10}As_3Mn$ molecule. Isosurfaces of the positive (green) and negative (orange) parts of the highest occupied and lowest unoccupied molecular orbits (HOMOs and LUMOs, respectively) corresponding to the isovalues 0.01 and 0.1 [smaller orbital surfaces in (f) and (i)]. (a) and (b): HOMO 118; (c) HOMO 119; (d) to (f): HOMO 120; (g) to (h): LUMO 122. Indium atoms are yellow, As red and Mn blue. Atomic dimensions are reduced to show the surface structure; other dimensions are to scale. In (b) and (i) MO surfaces are transparent to show contributing atoms.

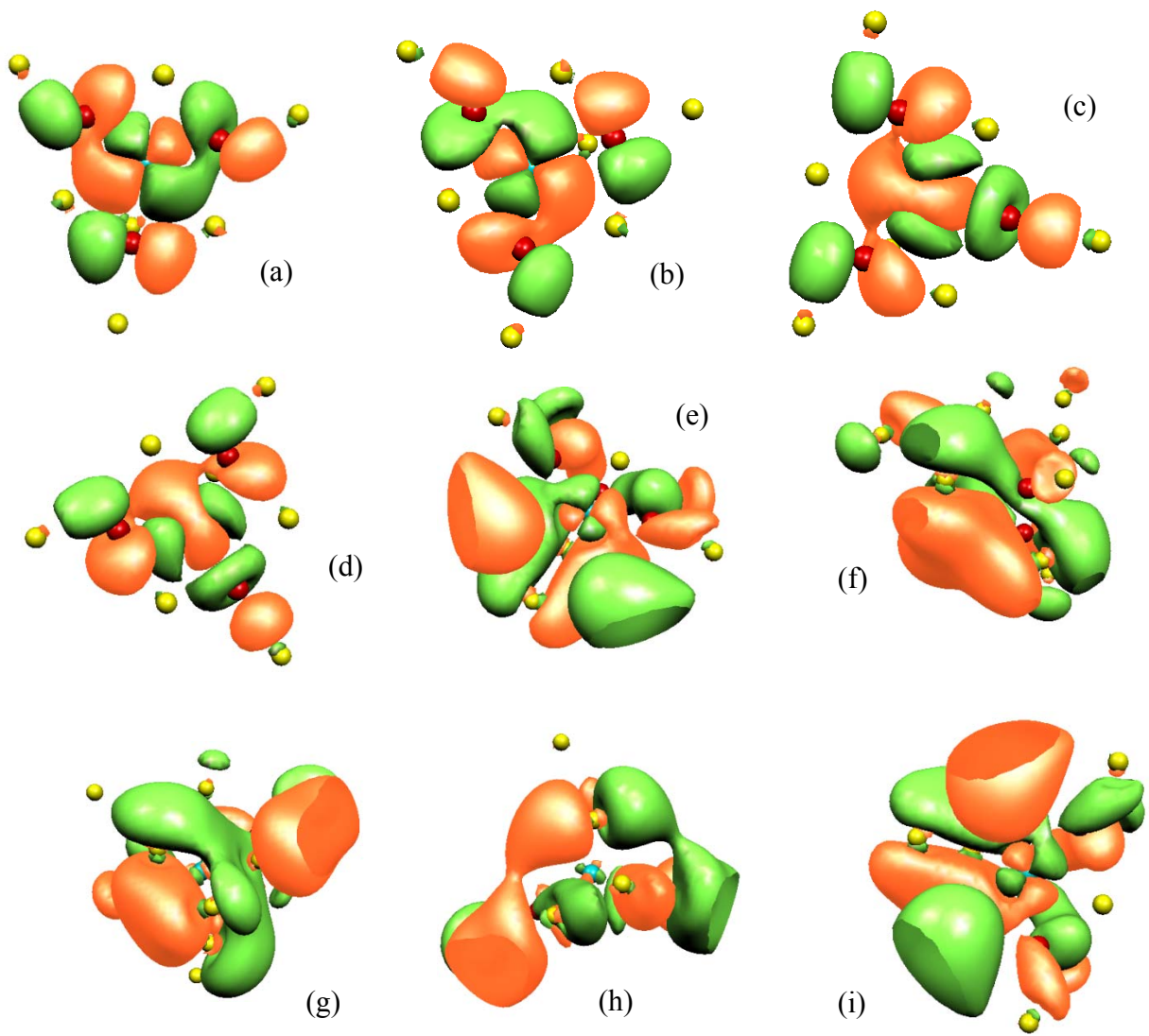

Fig. 8. (Color online) Vacuum In$_{10}$As$_3$Mn molecule. Isosurfaces of the positive (green) and negative (orange) parts of the highest occupied and lowest unoccupied molecular orbits (HOMOs and LUMOs, respectively) corresponding to several isovalues. (a) and (b): HOMO 121, isovalue 0.005; (c) and (d): HOMO 122, isovalue 0.005; (e) and (f): HOMO 123, isovalue 0.01; (g), (h) and (i): LUMO 124, 125 and 126, respectively, isovalue 0.015. In atoms are yellow, As red and Mn blue. Atomic dimensions are reduced to show the isosurface structure; other dimensions are to scale.

Several MOs in the HOMO – LUMO region of the vacuum molecule are depicted in Fig. 8. In this case, all MOs in the immediate HOMO – LUMO region are essentially shaped by hybridization of 3*d* AOs of the Mn atom and 4*p* AOs of As atoms, in agreement with suggestion of Ref. (79). The lower HOMO 121 (Figs. 8a and 8b) bonds In atoms through As atoms in two equal parts of the molecule, while there is no bonding between those two parts. The HOMO 122 (Figs. 8c and 8d) is a bonding MO with the major contributions from the $3d_z^2$ AO of the Mn atom and 4*p* AOs of As atoms, and very samll contributions from 4*d* AOs of In atoms. In contrast, the proper HOMO 123 (Figs. 8e and 8f) has significant contributions from 4*d* orbits of two In atoms, in addition to the major contributions from $3d_{xy}$ AOs of the Mn atom and 4*p* AOs of As atoms. Similar to HOMO 121 and 122, HOMO 123 is a bonding MO. The lower LUMOs 124, 125 and 126 (Figs. 8g to 8i) are also derived from the above mentioned AOs of In, As and Mn atoms, but these MOs feature significant ligand (In atoms) bonding contributions. In particular, the major contribution to LUMO 124 (Fig. 8g) comes from bonding of 4 In atoms between themselves through their 5*p* AOs (a "sandwich" in the lower left part of Fig. 8g that consists of a large negative "football" surface in the lower left part of Fig. 8g and the corresponding positive "lid" on the top), and with 3 other In atoms whose bonding is mediated by 3*d* AOs of Mn and 4*p* AOs of As atoms. The two higher MOs in the LUMO region, LUMO 125 and LUMO 126, are also bonding MOs, but ligand bonding there is realized through bonding with Mn and As atoms. The discussed MO properties of In10As3Mn molecules manifest violation of the octet rule of the standard valence theory, in agreement with other available observations (122, 119 – 121) and indicate that there may exist a wide range of non-stoichiometric molecules that are stable either on their own, or can be stabilized by their environment, such as quantum confinement.

    The HF and ROHF MOs are not very accurate, and the corresponding OTEs of the two In-based molecules are rather large (Table 2), which is typical for HF approximation (see a discussion in Chapter 2). However, the obtained results hint at the existing opportunities to

manipulate electronic and magnetic properties of small molecules composed of semiconductor compound atoms using the synthesis environment, such as quantum confinement, and adjusting the composition. The studied case demonstrates rich prospects opened by the use of quantum confinement as an element of molecular synthesis conditions. In particular, the pre-designed $In_{10}As_3Mn$ molecule virtually synthesized at conditions mimicking quantum confinement is a HF singlet ("antiferromagnetic" electron spin arrangement in the molecule) with the total magnetic moment equal to zero and HF OTE about 3.9 eV. At the same time, in the absence of spatial constraints (quantum confinement) the same atoms self-assemble into a different molecule with a relatively large magnetic moment and ROHF OTE about 1.24 eV.

While hardware restrictions do not allow virtual synthesis of much larger atomic systems (see Chapter 3 for more information), the discussed results in the case of small $In_{10}As_3Mn$ molecules lead to several important conclusions concerning electronic and magnetic properties of thin films of, and possibly bulk, InAsMn. [Note, that the composition of the studied molecules correspond to about 7% of Mn in the zincblende InAs structure, that is, to the upper limit of the case of "diluted magnetic semiconductors".] Based on the analysis of CDDs and SDDs of the $In_{10}As_3Mn$ molecules and those of $In_{10}As_4$ molecules analyzed in Chapter 4, one can conclude that electron charge deficit in the immediate vicinity of the "surface" and "inside" of the $In_{10}As_3Mn$ molecules (that is, regions of positive MEP somewhat below 1 nm in linear dimensions) is, to a degree, orchestrated by the Mn atom. In the language of the semi-phenomenological theory of semiconductors this charge deficit is called "hole". While average linear dimensions of a system are orders of magnitude larger than 1 nm (the case of bulk solids, thick films, large QDs and QWs, etc.) such a "hole" may be considered localized in the vicinity of a Mn atom. However, in the case of nanometer-thin films that are widely investigated at present (see Sec. I) such a hole cannot be treated as a localized object, because its "localization dimensions" are of the order of the thickness of the film. Therefore, in the case of thin films, small QDs and QWs of InAsMn with a few percent of Mn and characteristic dimensions of several nanometers, one must consider such a hole as a region of electron charge deficit delocalized over the outer "surface" of the pyramidal element of the InAs zincblende lattice composed of 10 to 14 In and As atoms containing a Mn atom as an As-substitution defect. Moreover, even in the pre-designed case (that is, the case reflecting conditions of the InAs zinkblende lattice) that region of electron charge deficit is not uniform, and is not centered on Mn atom. Instead, that region is composed of a system of sub-regions of electron charge deficit with centered near the geometrical center of the pyramids. This conclusion is further supported by data on the SDDs of the studied molecules (Fig. 5). In particular, for both molecules one can observe a large spin density values in the region including the pyramid geometrical center where there is no atoms in the InAs zincblende lattice.

Using the virtual synthesis data discussed above one can finally answer the question concerning "the place of residence" of the "holes". In particular, in the case when a nanometer-thick InAsMn film/system is sandwiched between some other films/systems (such systems are widely studied at present, see Sec. I) the holes do not reside exclusively inside of the InAsMn film/system. Rather, they include portions of the confining systems.

Further on, the ELS of the studied $In_{10}As_3Mn$ molecules differs significantly, especially in the HOMO-LUMO regions. This means, that for systems with one or more linear dimensions in the range of a few nanometers, the "band" structure (if such a structure still can be properly identified) is not derived directly from InAs or Mn "valence bands". Instead, it should be treated as a new band structure originating from quantum confinement and Coulomb effects governing quantum motion of strongly correlated electrons in broken symmetry systems. Thus, reasoning about the "impurity band" and its position related to the "valence band" of the host lattice is meaningless for small systems with one or more linear dimensions in the range of a few nanometers.

## 4. PRE-DESIGNED AND VACUUM MOLECULES $In_{10}As_3V$

Virtual synthesis studies of $In_{10}As_3V$ molecules are of significant importance both for practical purposes and for understanding a role of $3d$ AOs of V and Mn atoms in the development of positive MEP regions (delocalized holes) in InAs and GaAs zincblende bulk lattices containing a few percent of Mn or V atoms. Similar to the pre-designed $In_{10}As_3Mn$ molecule, the pre-designed $In_{10}As_3V$ molecule has been developed by substitution of V atom instead of As one in the pyramidal symmetry element of the zincblende InAs lattice, and subsequent HF/ROHF minimization of the total energy of the atomic cluster so obtained in the case when spatial constraints were applied to the centers of mass of the cluster's atoms. The corresponding vacuum molecule has been obtained upon the total energy minimization of the pre-designed cluster in the case when the spatial constraints applied to the centers of mass of its atoms were lifted, so the atoms could move. The structure of the molecules so obtained is detailed in Fig. 9. Visually these two molecules are indistinguishable, although in the vacuum molecule several atoms, including V and all As atoms have moved by several tens of Angstrom from their original positions in the pre-designed molecule. This motion also resulted in small changes to the angles between V, In and As atoms. For example, in the case of tetrafold-coordinated As atoms in the pre-designed molecule the In-As-In angles with the closest 4 In atoms are 113.0º, 113.0º, 105.7º and 105.7º, while in the case of the vacuum molecule those angles are 118.7º, 110.6º, 105.6º and 99.7º for two of the As atoms on the zincblende cube body diagonal opposite to that of the V atom, and 111.8º, 111.1º, 107.7º, and 107.0º for the As atom sharing the zincblende cube body diagonal with V atom. The vanadium atom has also been displaced from its original position in the pre-designed molecule. Thus, the In-V-In angles in the small, tetrafold-coordinated V pyramid of the vacuum molecule are 118.4º, 111.0º, 105.4º and 99.6º. These small adjustments, however, are such that they do not affect V - As atomic coordination: both the distances between V and As atoms, and the angles remain the same as they are in the pre-designed molecule – 3.997 Å and 60.5º, respectively.

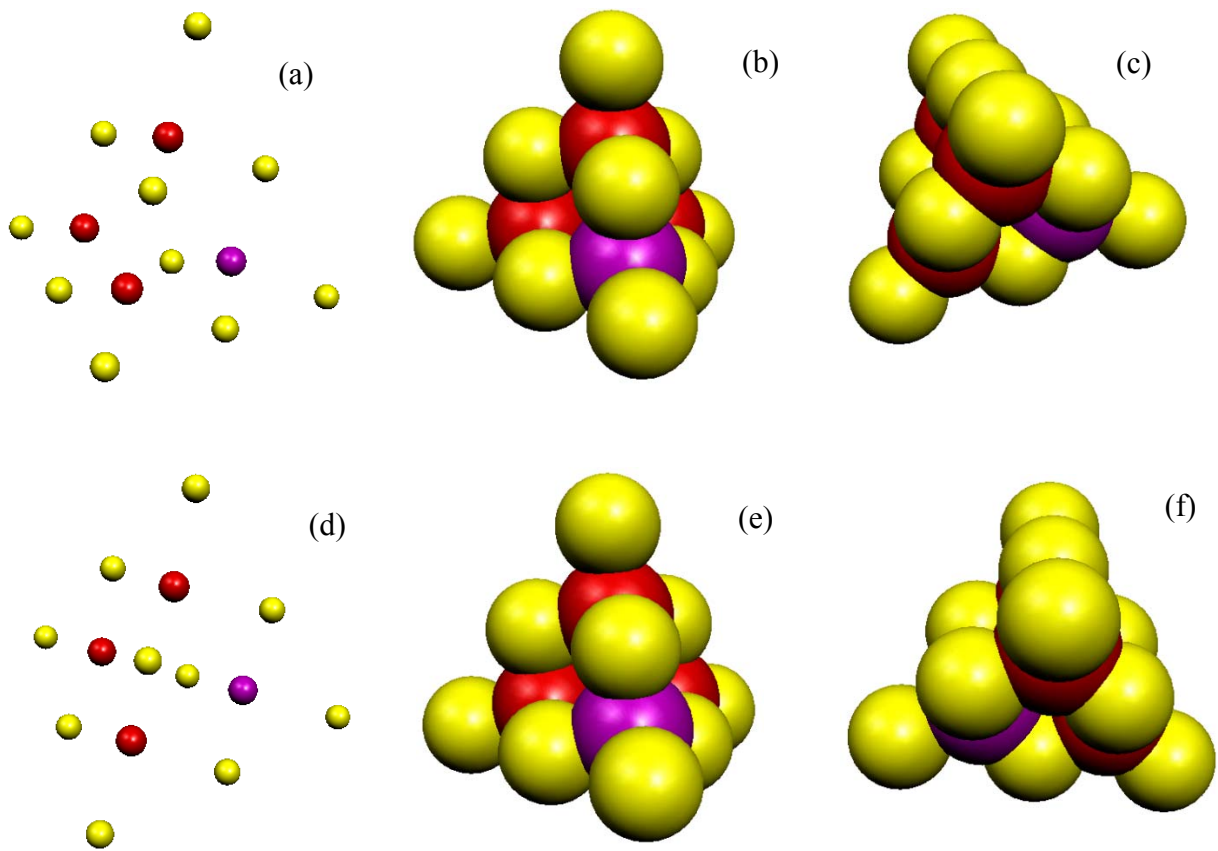

Fig. 9. (Color online) The structure of the pre-designed [(a) to (c)] and vacuum [(d) to (f)] $In_{10}As_3V$ molecules. Indium atoms are yellow, As red and V purple. In (a) and (d) atomic dimensions are reduced to show atomic positions; other dimensions are to scale. The radii of In atoms in (b), (c), (e) and (f) are somewhat smaller than their covalent radius, and those of As and V atoms are somewhat larger than the corresponding covalent radii of As and V atoms.

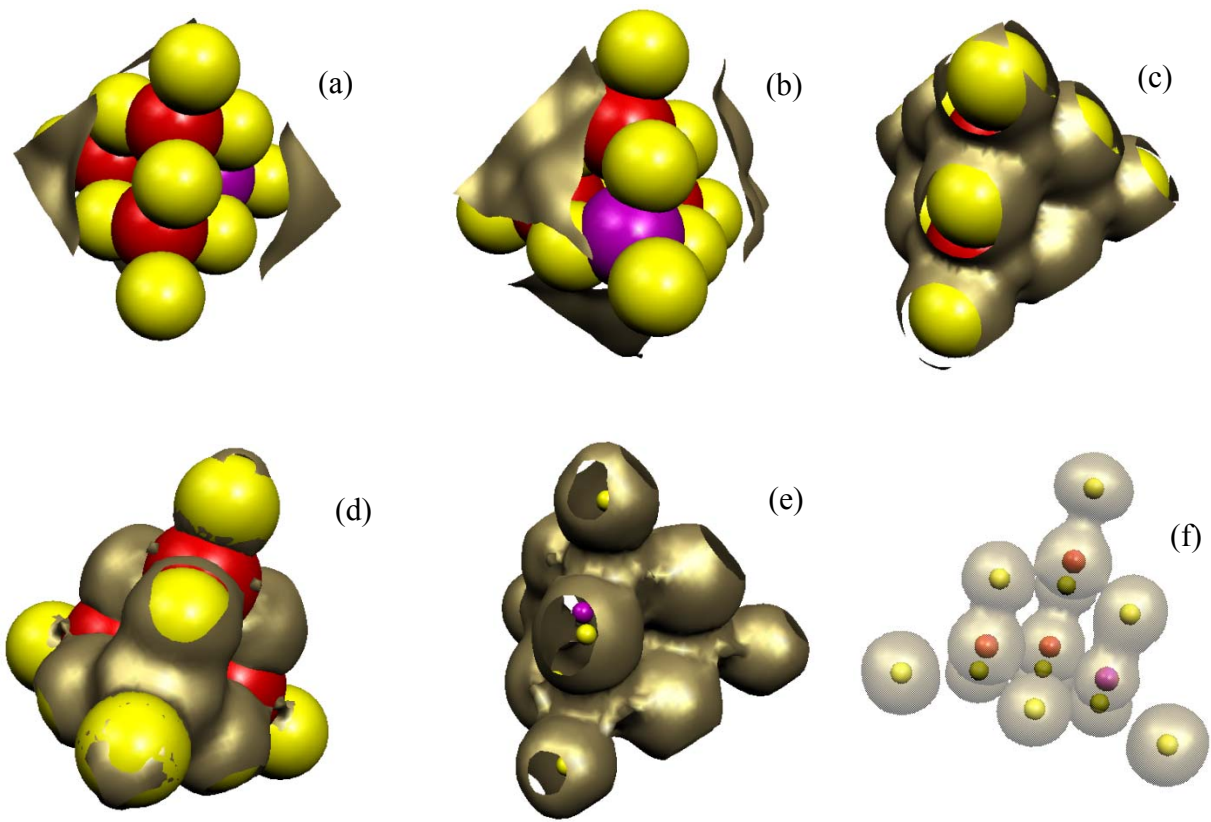

Fig. 10. (Color online) The electron charge density distribution (CDD) of the pre-designed $In_{10}As_3V$ molecule. Isosurfaces (golden) corresponds to the isovalues (a) 0.0005; (b) 0.01; (c) 0.05; (d) 0.075, (e) 0.075, and (f) 0.15. Indium atoms are yellow, As red and V purple. In (e) and (f) atomic dimensions are reduced to show the isosurface structure; other dimensions are to scale. The radii of In atoms in (a) to (d) are somewhat smaller than their covalent radius, and those of As and V atoms are somewhat larger than the corresponding covalent radii of As and V atoms.

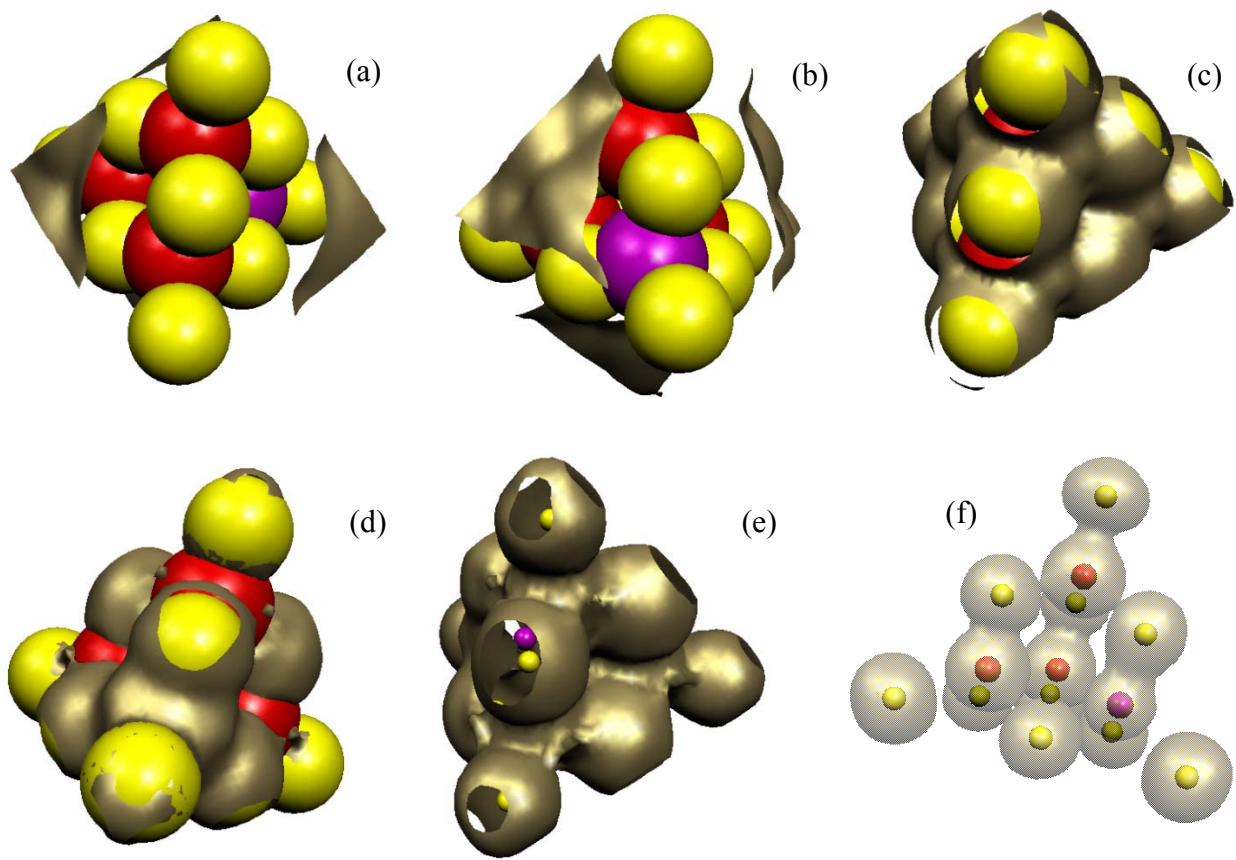

Fig. 11. (Color online) The electron charge density distribution (CDD) of the vacuum $In_{10}As_3V$ molecule. Isosurfaces (golden) corresponds to the isovalues (a) 0.001; (b) 0.01; (c) 0.05; (d) 0.05, (e) 0.15, and (f) 0.25. Indium atoms are yellow, As red and V purple. In (e) and (f) atomic dimensions are reduced to show the isosurface structure; other dimensions are to scale. The radii of In atoms in (a) to (d) are somewhat smaller than their covalent radius, and those of As and V atoms are somewhat larger than the corresponding covalent radii of As and V atoms.

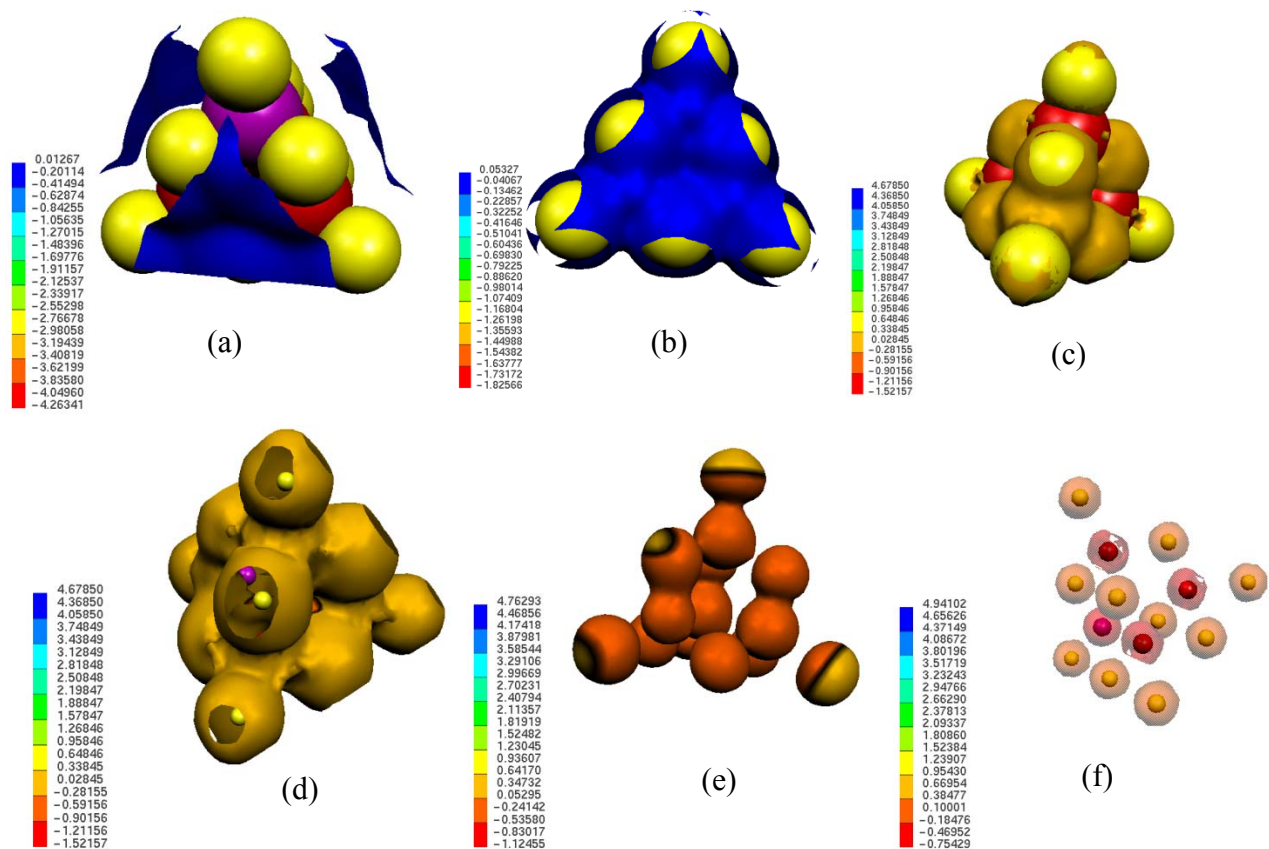

Fig. 12. (Color online) The molecular electrostatic potential (MEP) of the pre-designed molecule In$_{10}$As$_3$V for several isosurfaces of the CDD calculated for the following fractions (isovalues) of the CDD maximum value 3.54328 (arbitrary units): (a) 0.01; (b) 0.05; (c) and (d) 0.075; (e) 0.15, and (f) 0.3. The color coding scheme for MEP surfaces is shown in each figure. In atoms are yellow, As red and V purple. All dimensions in (a) to (d) are to scale; atomic dimensions roughly correspond to the atoms' covalent radii. In (d) to (f) atomic dimensions are reduced to show the MEP surface structure.

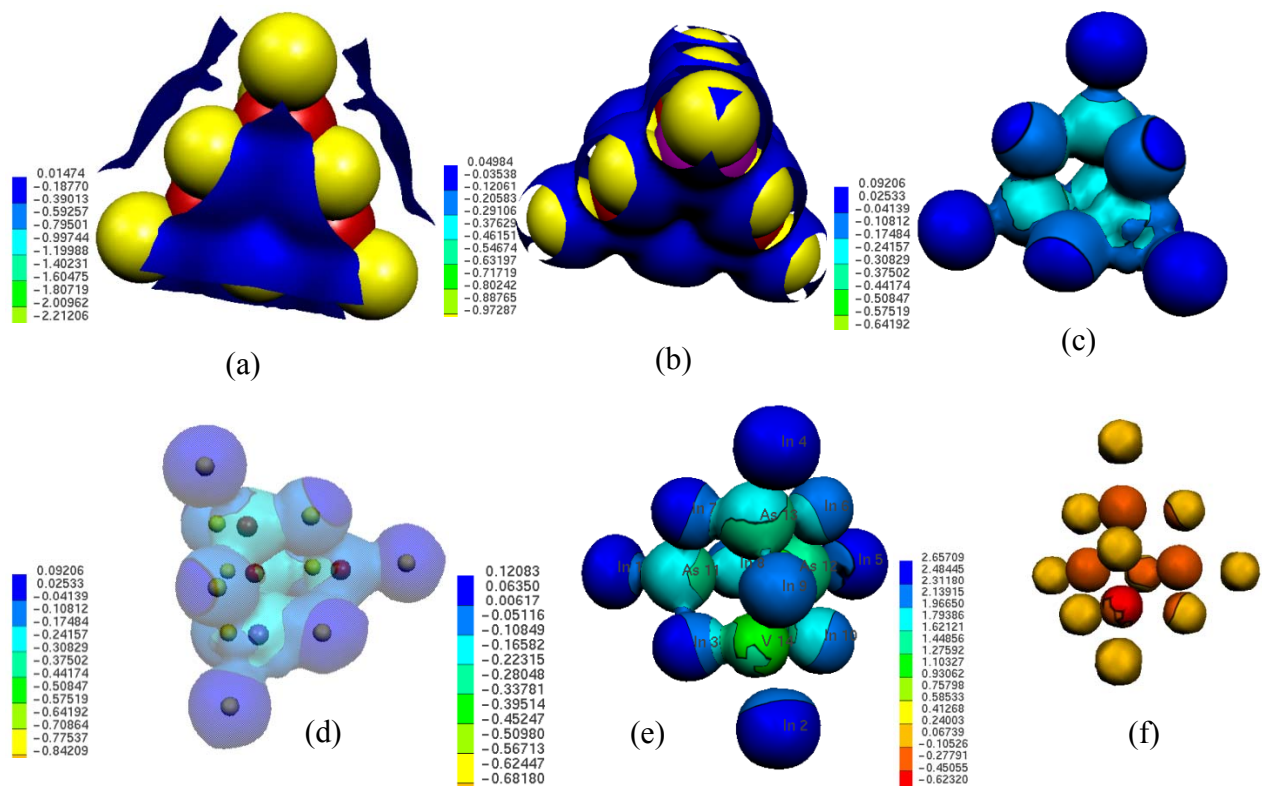

Fig. 13. (Color online) The molecular electrostatic potential (MEP) of the vacuum molecule $In_{10}As_3V$ for several isosurfaces of the CDD calculated for the following fractions (isovalues) of the CDD maximum value 3.264001 (arbitrary units): (a) 0.01; (b) 0.05; (c) and (d) 0.1; (e) 0.15, and (f) 0.25. The color coding scheme for MEP surfaces is shown in each figure. In atoms are yellow, As red and V purple. All dimensions in (a) and (b) are to scale; atomic dimensions roughly correspond to the atoms' covalent radii. In (c) to (f) atomic dimensions are reduced to show the MEP surface structure.

CDDs (Figs. 10 and 11) and MEPs (Figs. 12 and 13) of both molecules retain a general appearance of tetrahedral symmetry. However, detailed analysis reveals tetrahedral symmetry breaking in both cases. Comparing these CDD and MEP isosurfaces to those of $In_{10}As_3Mn$ molecules one can see that in the latter case the electron charge is pushed further into the space surrounding the molecules with Mn atom. In the case of molecules containing V atom the electronic charge are about twice as closer to the molecular "surfaces" for the same isosurface values. The vanadium atoms in these molecules accumulate electronic charge (Figs. 10d and 11d). The MEP values are positive near the "surfaces" of both molecules (Figs. 12d and 13d) and in the immediate vicinity of the "surfaces" on the inner side of the molecules (Figs. 12 e to 12f, and 13e to 13f). The "shell" of positive MEP values surrounding molecular surfaces is less distinctive and the MEP values are lesser in the case of the pre-designed $In_{10}As_3V$ molecule (Figs. 10 and 12) than those in the case of its vacuum counterpart. To a degree, this may be a consequence of the fact that the pre-designed molecule is a ROHF nonet (its spin multiplicity M is 9). Such high excited states are beyond applicability of the ROHF approximation (see the corresponding discussion in Chapter 3), indicating that this molecule may not be stable. In contrast, the vacuum vanadium-containing molecule is a ROHF pentet, and therefore is expected to be much more stable. Correspondingly, its "shell" of positive MEP values is much more distinctly defined and contains regions in the immediate outer vicinity of the molecular "surface", and much thicker regions near that "surface" from inside of the molecule. Using the "hole" terminology, one can expect the hole mediated by V atom to be contained inside the 14-atomic tetrahedral symmetry element of the zincblende InAs structure containing a vanadium substitution defect, while in the case of Mn-mediated "holes" of $In_{10}As_3Mn$ molecules such "holes" are delocalized about a larger region containing the corresponding pyramidal element. The volume of that region is about twice as large as the region of delocalization of V-mediated "holes" in $In_{10}As_3V$ molecules. The observation that the V-mediated charge hole delocalized in a pyramidal symmetry element of the InAs zincblende lattice containing the vanadium atom is more localized than the corresponding Mn-mediated hole steams from the fact that a vanadium atom has smaller number of 3$d$-electrons (2, as opposed to 5 3$d$-electrons of a Mn atom), and those electrons are closer to the V nucleus. Therefore, more ligand electron charge of In atoms can be kept "inside" of $In_{10}As_3V$ molecules (Figs. 12e and 13e) compared to the case of $In_{10}As_3Mn$ molecules (Figs. 3e and 4e). In practical terms this means that the semi-phenomenological band theory of semiconductors and its extensions and modifications designed to embrace a realm of DMS should work better for InAsV systems than for InAsMn ones.

Analysis of MOs of $In_{10}As_3V$ molecules provides detailed information on electron charge configuration, a role of the vanadium atom in stabilization of these molecules and the development of regions of positive MEP values ("holes"). Several such MOs in the HOMO-LUMO regions of these molecules are depicted in Figs. 14 and 15. Similar to the case of $In_{10}As_3Mn$ molecules, the major contributions to MOs of $In_{10}As_3V$ molecules in the HOMO-LUMO region come from $5p$ AOs of In atoms, $3d$ AOs of V atom and $4p$ AOs of As atoms hybridized in various proportions, and some contributions from $4d$ AOs of In atoms (this latter contributions must be ascertained by further CI, MCSCF and MP-2 studies). In the case of the pre-designed ROHF nonet, HOMO 121 (Figs. 14a and 14b) contains both bonding and non-bonding regions mediated by $3d_{xy}$ AOs of V atom that orchestrates In ligand bonding via their $5p$ AOs, with a significant contribution of $4p$ AOs of As atoms. HOMOs 122 and 123 are bonding MOs with the major contributions from $3d_{z2}$ AOs of the vanadium atom hybridized with both $5p$ and $4d$ AOs of several In ligand atoms (Figs. 14c to 14f). Only $4p$ AOs of one As atom contribute to bonding in the case of HOMO 122 (Fig. 14d), while all As atoms significantly contribute to bonding in the case of HOMO 123 (Figs. 14e and 14f). LUMOs 124 and 126 of this molecule (Figs, 14g and 14i, respectively) are bonding MOs where the In ligand bonding is mediated by $3d_{z2}$ AOs of the vanadium atom, while LUMO 125 contains both non-bonding and bonding regions all mediated by $3d_{xy}$ AOs of the vanadium atom. In contrast to $In_{10}As_3Mn$ molecules, all MOs in the HOMO-LUMO region of the pre-designed $In_{10}As_3V$ molecule exhibit significant contributions from ligand bonding with 4 or more participating In atoms. Both $5p$ some $4d$ AOs of In atoms contribute to this bonding, in agreement with the corresponding CDD picture indicating that in the case of V-containing molecules electron charge of In atoms is more effectively redistributed inside the molecules, and less of this charge is pushed outside of the molecular "surfaces".

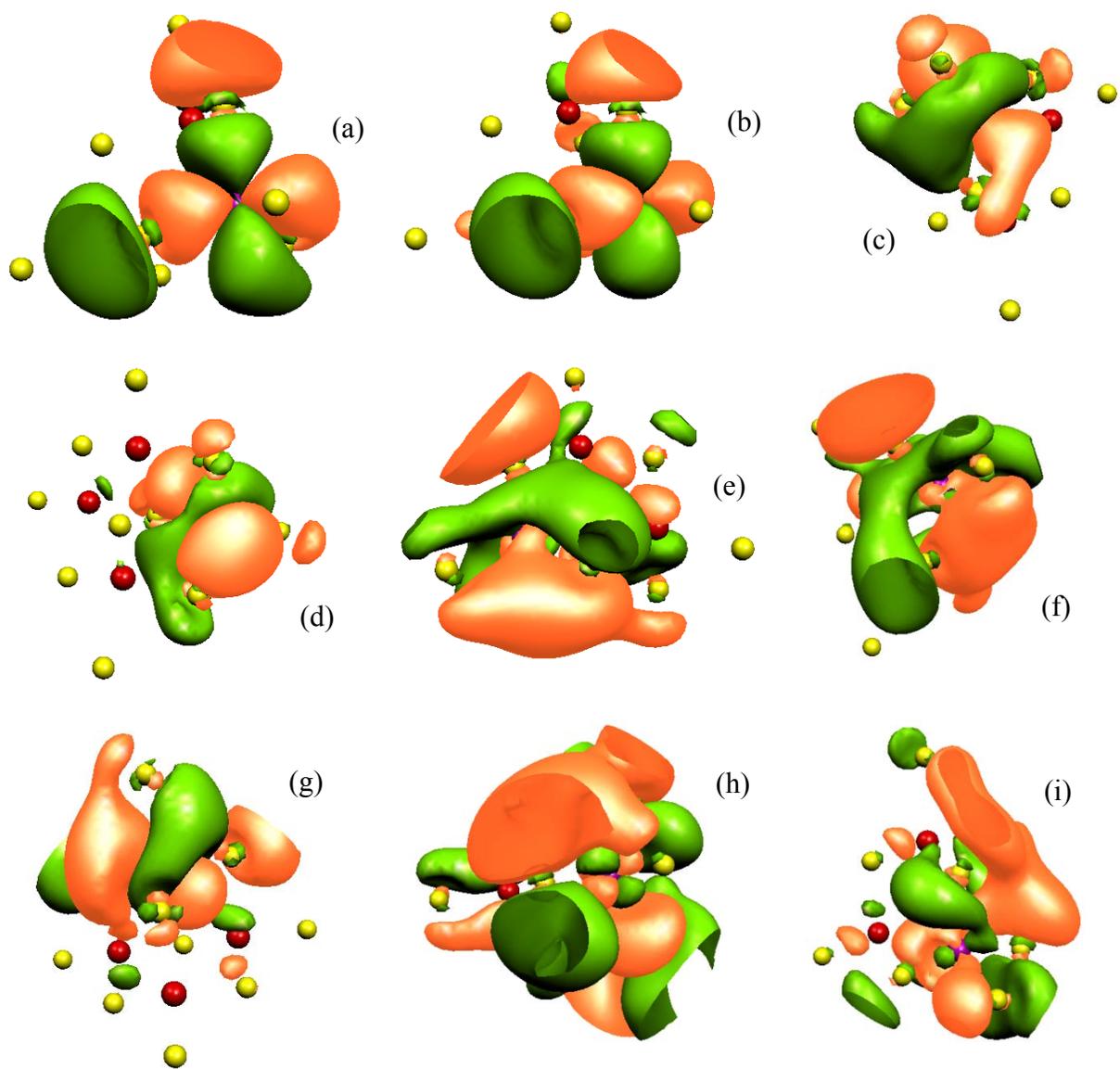

Fig. 14. (Color online) The pre-designed $In_{10}As_3V$ molecule. Isosurfaces of the positive (green) and negative (orange) parts of the highest occupied and lowest unoccupied molecular orbits (HOMOs and LUMOs, respectively) corresponding to several isovalues. (a) and (b): HOMO 121, isovalue 0.01; (c) and (d): HOMO 122, isovalue 0.01; (e) and (f): HOMO 123, isovalue 0.015; (g), (h) and (i): LUMO 124, 125 and 126, isovalues 0.0150.01, 0.01 and 0.02, respectively. Indium atoms are yellow, As red and V purple. Atomic dimensions are reduced to show the isosurface structure; other dimensions are to scale.

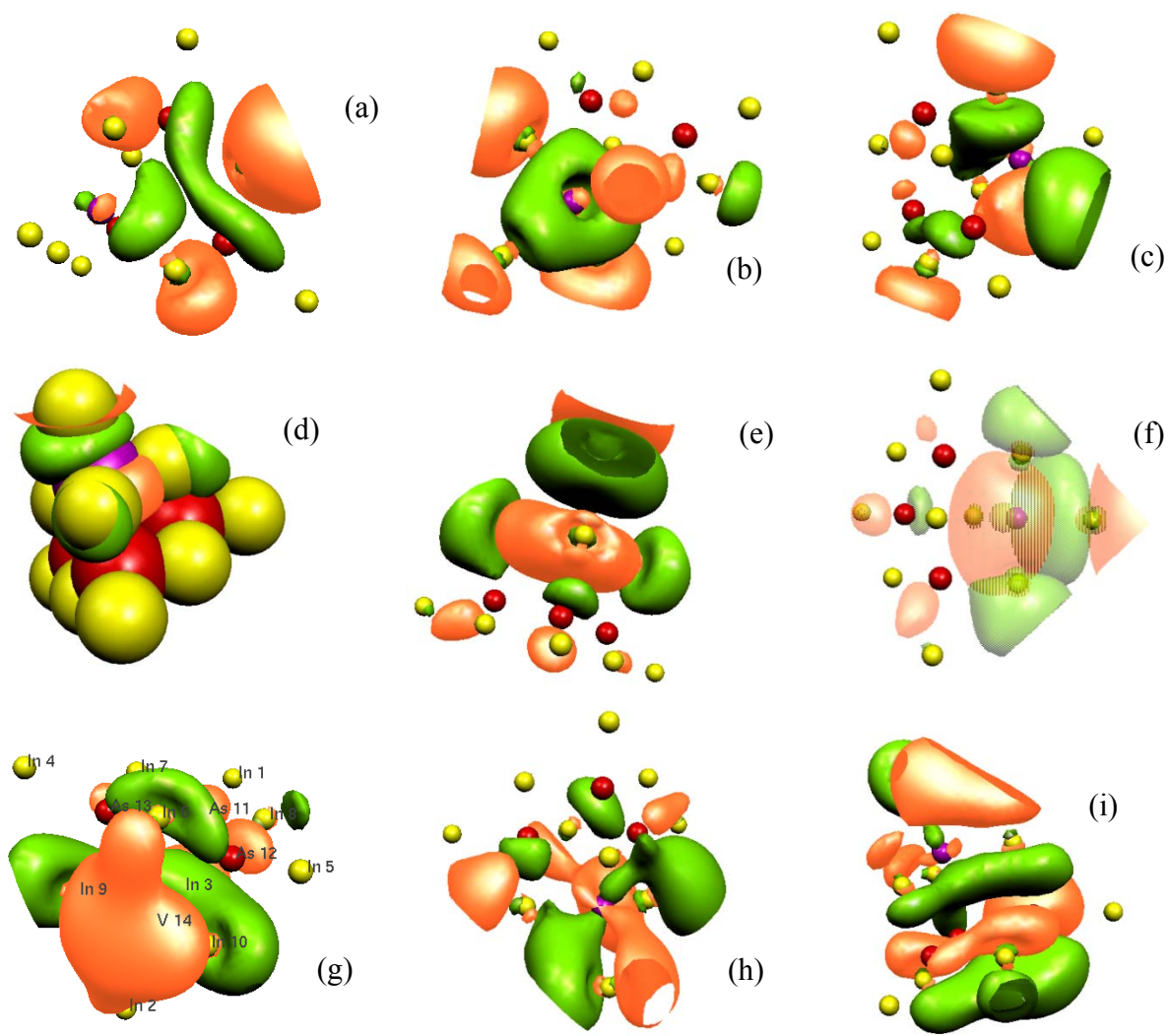

Fig. 15. (Color online) The vacuum $In_{10}As_3V$ molecule. Isosurfaces of the positive (green) and negative (orange) parts of the highest occupied and lowest unoccupied molecular orbits (HOMOs and LUMOs, respectively) corresponding to several isovalues. (a) to (c): HOMO 118, 119 and 120, isovalue 0.01, respectively; (d) and (f): HOMO 121, isovalue 0.01; (g) and (h): LUMO 122 and 123, isovalue 0.015, respectively; and (i) LUMO 124, isovalue 0.015. Indium atoms are yellow, As red and V purple. In all figures except for (d) atomic dimensions are reduced to show the isosurface structure; other dimensions are to scale. In (d) atomic dimensions are roughly defined by the atoms' covalent radii.

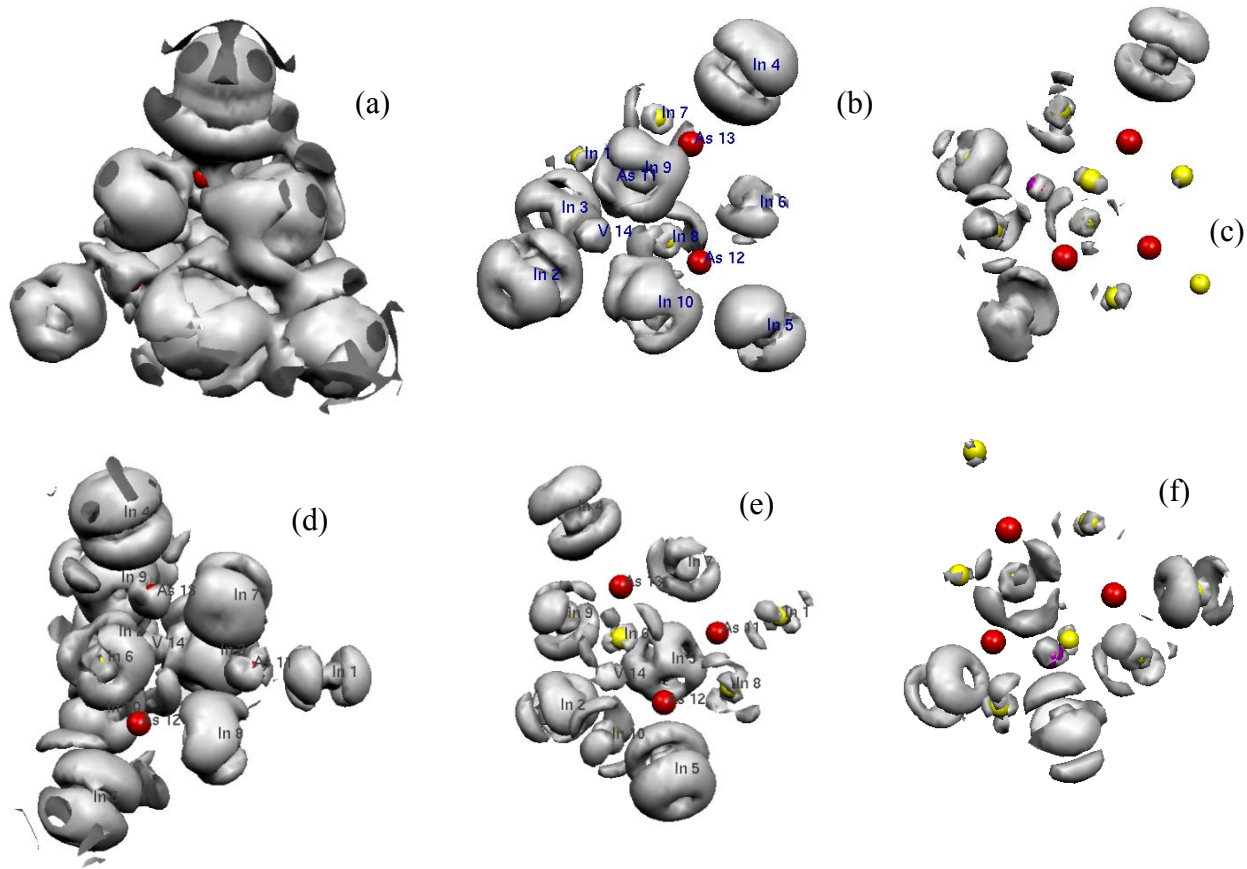

Fig. 16. Isosurfaces of the spin density distribution (SDD) of the pre-designed [(a) to (c)] and vacuum [(d) to (f)] $In_{10}As_3V$ molecules corresponding to the fractions (a) 0.001, (b) 0.005, (c) 0.01, and (d) 0.001, (e) 0.003, (f) 0.005 of the respective SDD maximum values (not shown). Indium atoms are yellow, As red and Mn blue. All atomic dimensions are reduced to show the SDD surface structure.

All MOs in the HOMO-LUMO region of the vacuum $In_{10}As_3V$ molecule shown in Fig. 15 are bonding orbits. The $3d_{z2}$ AO of the vanadium atom provide the major contribution to all MOs except for LUMO 124, in which case the contribution due to this AO of the vanadium atom are small and all bonding is mediated by $4p$ AOs of 3 As atoms. All MOs of this molecule exhibit two large areas of shared charge: one is a delocalized charge of 4 to 6 In ligand atoms mediated by V, and the other a delocalized charge of 3 In atoms mediated by one to three As atoms. These two portions of the MO are bonded to each other through one of the In ligand atoms. Once again, this type of bonding violates the octet rule and is the major means of stabilization of the studied non-stoichiometric molecules.

The In ligand bond length in the case of the pre-designed molecule is defined entirely by the geometry and is 4.284 Å, which is also the distance between any two In neighbours. In the vacuum molecule the ligand bond length is flexible taking several values between 4.272 Å and 4.999 Å. The In-As bond length is flexible in both cases and allows for several values in the ranges from 2.220 Å to 2.934 Å (the pre-designed molecule) and 2.346 Å to 2.895 Å (the vacuum molecule). The arsenic and vanadium atoms bond directly only to In ones. In the case of the pre-designed molecule the length of the V - In bond can take 4 values: 2.220 Å, 2.717 Å, 2.934 Å and 4.695 Å. The first 3 values are the same as those of the In - As bond in this molecule. In the case of the vacuum molecule the length of V - In bond takes only 3 values: 2.340 Å, 2.706 Å and 2.907 Å.

Both $In_{10}As_3V$ molecules are ROHF spin multiplets with "ferromagnetic" arrangement of uncompensated spins (Fig. 16). The spin of these molecules are primarily contributed to by spin components of delocalized $4d$ electrons of In atoms, with a very small contribution coming from electrons of the V atom. AOs of arsenic atoms do not contribute to SDDs. SDD values of the pre-designed molecule (which is a "ferromagnetic" ROHF nonet with the uncompensated magnetic moment $9\mu_B$, Figs. 16a to 16c) are about 3 times larger than those of the vacuum one, which is a "ferromagnetic" ROHF pentet with the uncompensated magnetic moment $5\mu_B$, Figs. 16d to 16i. Thus, the InAs-based molecules with one vanadium atom are stronger "magnets", and thus more suitable for DMS applications. In particular, the pre-designed $In_{10}As_3V$ molecule is "ferromagnetic" and possesses the largest magnetic moment among the studied InAs-based molecules. At the same time, the pre-designed $In_{10}As_3Mn$ molecule is "antiferromagnetic" singlet with its zero uncompensated magnetic moment. [The latter finding is consistent with experimental observation that with a change in thermodynamic conditions some thin DMS films exhibit magnetic phase transitions; see Sec. 1 for further details and references.] At the same time, much larger and heavier "holes" mediated by Mn atoms in $In_{10}As_3Mn$ structures may have their own use for applications.

## 5. $Ga_{10}As_3V$ MOLECULES WITH ONE VANADIUM ATOM.

Interest to GaAs-based DMS is rising, because such systems have some technological advantages over InAs-based DMS systems, and may be simpler to understand than GaAsMn-type of DMS. At present, GaAsV DMS systems have not been well investigated, so virtual synthesis and computational studies of basic GaAsV structures may provide guidance to experimentalists and engineers.

Similar to the studied $In_{10}As_3V$ molecules, two $Ga_{10}As_3V$ molecules have been virtually synthesized using computational procedures discussed in Sec.2 and Chapter 3. Thus, the pre-designed $Ga_{10}As_3V$ molecule was obtained by the total energy minimization procedure applied to a tetrahedral symmetry element (a pyramid) of the zincblende GaAs lattice described in Chapter 3, where one of As atoms was replaced by a vanadium one. During such conditional energy minimization all positions of the centers of mass of the pyramid atoms were kept fixed. The corresponding vacuum $Ga_{10}As_3V$ molecule was virtually synthesized by lifting the special constraints applied to the centers of mass of the atoms in the pre-designed molecule, and minimizing the total energy of the atomic cluster unconditionally (that is, without any constrains applied).

The structure of the obtained molecules is depicted in Fig. 17. To a human eye, these structures seem to be the same, but analysis reveals that many atoms in the vacuum pyramidal molecule moved from their former positions in the pre-designed one. The pre-designed pyramid consists of 4 smaller pyramids built on As and V atoms. All distances between Ga and As atoms in the three As-coordinated pyramids are equal to 2.448 Å, and all Ga-As-Ga angles are 109.5º. This, of course, corresponds to a separation between an As atom and its 4 closest Ga neighbours, and related angles, respectively, in the GaAs zincblende lattice. The vanadium atom in this molecule simply substitutes an As one, so all dimension of V-driven small pyramid are equal to those of the small As-coordinated pyramids. All closest neighbor distances between As an V atoms are 3,997Å, and the corresponding angles 60º. Thus, geometrically, the pre-designed $Ga_{10}As_3V$ molecule is the perfect pyramid.

The vacuum molecule is far from being of perfect pyramidal structure. All As and V atoms in this molecule moved from their original positions corresponding to those in the perfect pre-designed pyramid. Thus, the distances and angles in the small As-topped pyramidal arrangements changed by several tenths of Å and from about 2º to 7º, respectively. In particular, the distances between Ga and V atoms in the V-topped small pyramid have become 2.505Å, 2.998Å, 2.998Å and 2.998Å, and the Ga-V-Ga angles 118.9º, 111.5º, 111.5º and 101.3º, respectively. The Ga-topped small pyramids have changed even more dramatically: the distances between Ga and As atoms in each of the pyramids have become 2.835Å, 2.519Å, 2835Å and 2.487Å. The sets of Ga-As-Ga angles in the As-coordinated small pyramids differ for each of the pyramids. For the pyramid coordinated by the As(13) atom (see the atomic numbering in Figs. 17c and 17f) the set of angles includes 58.5º, 58.5º, 103.7º and 107.9º; for the As(14)-coordinated pyramid the set includes 103.7º, 103,7º, 116.6º and 116.6º, and for the As(12)-topped pyramid it is 103.7º, 106.7º, 116.6º and 107.9º. This tetrahedral symmetry breaking has developed to stabilize a molecule when spatial constraints were applied to positions of its atoms were lifted.

The dipole moment of the perfect pre-designed pyramid $Ga_{10}As_3V$ is 1.387208 D. It is applied directly to the center of the pyramid base farthest from the V atom, and runs strictly along the pyramid heights toward the vertex Ga atom (Fig. 17c). In the case of the vacuum molecule the dipole moment is about 3 times larger: 4.569266 D, and is applied to the pyramid base closest to the V atom running through the V atom itself (Figs. 17e and 17f).

The MEPs of these molecules are pictured in Figs. 18 and 19. Similar to $In_{10}As_3Mn$ and $In_{10}As_3V$ molecules of Sec. 3 and 4, CDD and MEP surfaces of both $Ga_{10}As_3V$ molecules retain appearance of somewhat broken tetrahedral symmetry. Characteristic features of MEP surfaces of the pre-designed $Ga_{10}As_3V$ are close to those of $In_{10}As_3Mn$ molecules, while such features in the case of the vacuum $Ga_{10}As_3V$ molecule remind those of $In_{10}As_3V$ molecule. Indeed, the electron charge of the pre-designed molecule is pushed further outside of the molecule's "surface" (Figs. 18a to 18c), and also deeper inside of the molecule (Figs. 18d to 18f), so the "shell" of electron charge deficit surrounding the "surface" is much thicker than that of the vacuum $Ga_{10}As_3V$ molecule. The electron charge of the latter molecule is distributed relatively close to the molecular "surfaces" (Figs. 19a and 19b) on the outer side and inside of the molecular volume creating a relatively thin "shell" of electron charge deficit surrounding the molecular "surface" (Fig. 19a). In the case of this molecule the thickness of the electron charge deficit "shell" is about 2 covalent radii of Ga atom. The major reason for this striking difference in the electron charge distributions of the $Ga_{10}As_3V$ molecules is that the pre-designed molecule is strained. Indeed, the covalent radius of a Ga atom is much smaller than that of the In atom (Table I), and is close to that of the As atom. Thus, the partial "volume" occupied by V atom in the pre-designed $Ga_{10}As_3V$ molecule is larger than that in the case of the pre-designed $In_{10}As_3V$ molecule. As a result, replacement of an As atom by a V one causes much more electron charge imbalance in the pre-designed $Ga_{10}As_3V$ molecule as compared to that of the pre-designed $In_{10}As_3V$ molecule, and therefore, the former molecule is more strained than the latter one. One can conclude that the CDDs of the pre-designed and vacuum $In_{10}As_3V$ molecules should differ less between themselves than the CDDs of the pre-designed and vacuum $Ga_{10}As_3V$ molecules. This is confirmed by the obtained data illustrated in Figs. 18 and 19.

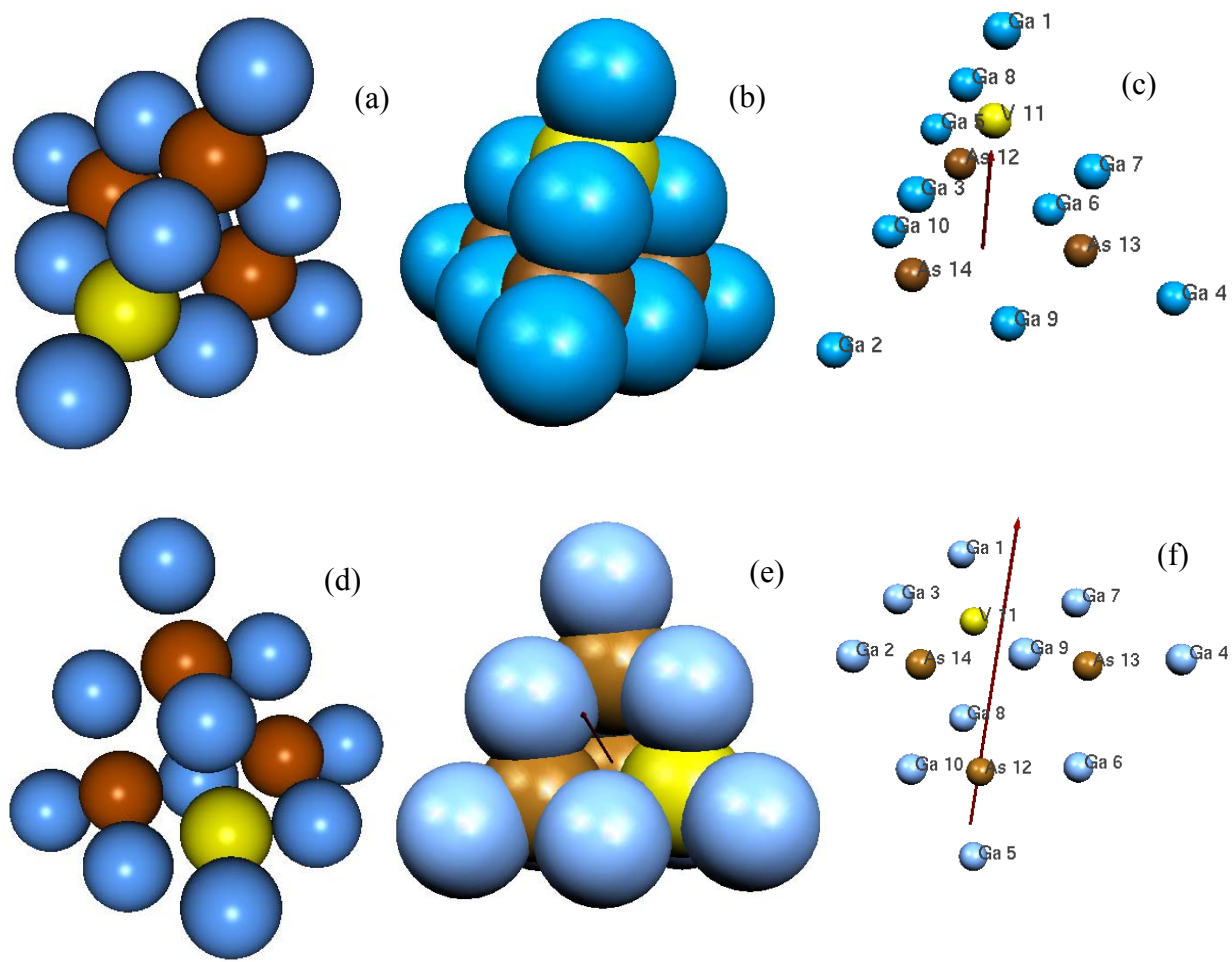

Fig. 17. (Color online) Pre-designed [(a) to (c)] and vacuum [(d) to (f)] molecules In$_{10}$As$_3$V. In (a) and (d) atomic dimensions approximately correspond to the atoms' covalent radii. In (b) and (e) atomic dimensions are enlarged to reveal the shape of the structures, and in (c) and (f) the atomic dimensions are reduced to show the dipole moment [red arrow in (c), (e) and (f)]. Gallium atoms are blue, As brown, and V yellow.

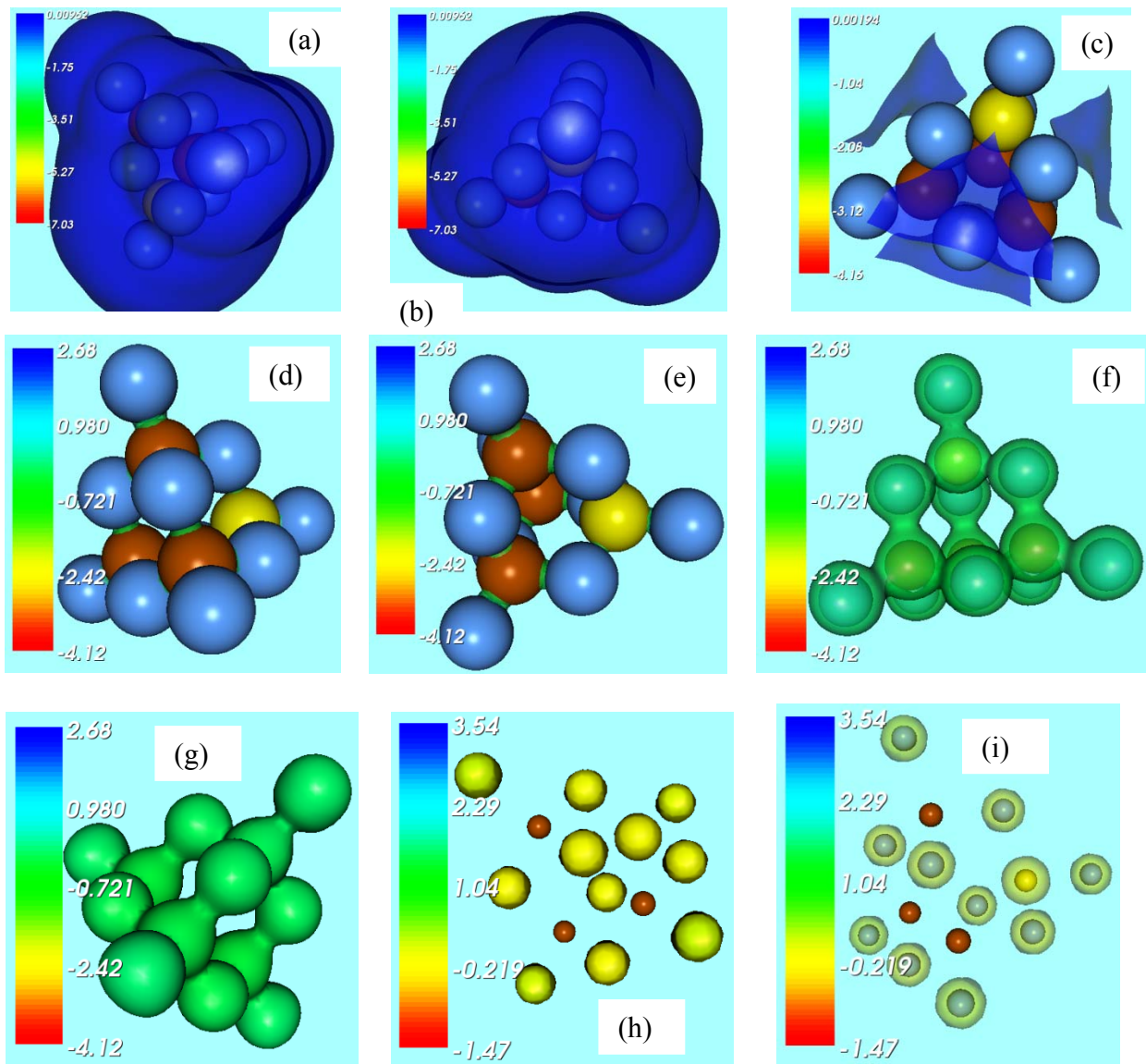

Fig. 18. (Color online) The molecular electrostatic potential (MEP) of the pre-designed molecule $Ga_{10}As_3V$ for several isosurfaces of the CDD calculated for the following fractions (isovalues) of the CDD maximum value (not shown). (a) and (b): 0.001; (c) 0.01; (d) to (g): 0.1; (h) and (i): 0.3. The color coding scheme for MEP surfaces is shown in each figure. Ga atoms are blue, As brown and V yellow. In (a) to (e) atomic dimensions are slightly smaller than those defined by the atoms' covalent radii, and in (f) to (i) atomic dimensions are significantly reduced to show the MEP surface structure. In (a) to (f) and (i) MEP surfaces are semi-transparent to reveal the structure.

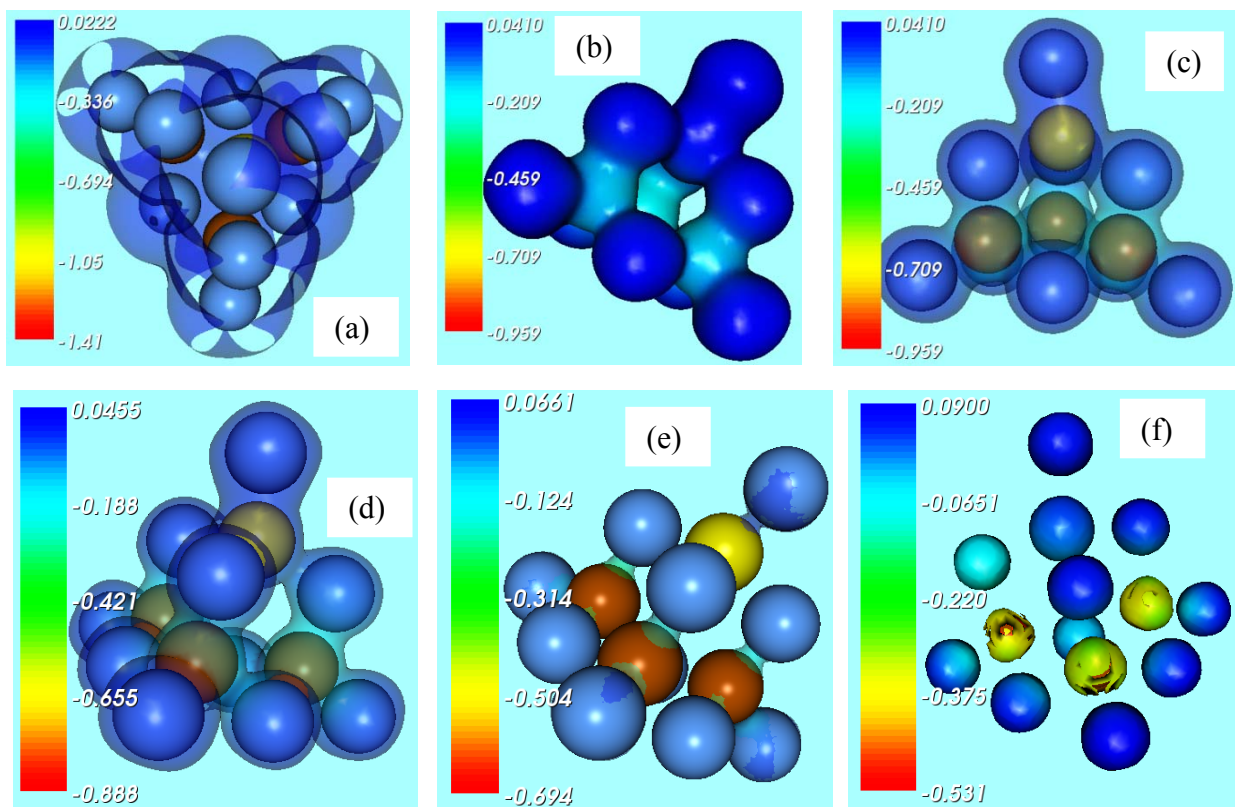

Fig. 19. (Color online) The molecular electrostatic potential (MEP) of the pre-designed molecule $Ga_{10}As_3V$ for several isosurfaces of the CDD calculated for the following fractions (isovalues) of the CDD maximum value (not shown): (a) 0.02; (b) and (c) 0.05; (d) 0.06; (e) 0.1, and (f) 0.15. The color coding scheme for MEP surfaces is shown in each figure. Ga atoms are blue, As brown and V yellow. In (a) to (e) atomic dimensions are somewhat smaller than those defined by the atom's covalent radii, and in (f) atomic dimensions are significantly reduced to show the MEP surface structure. In (a), (c), (d) and (e) MEP surfaces are semi-transparent to reveal the atoms.

There is yet another fact confirming the above conclusion that the pre-designed $Ga_{10}As_3V$ molecule is more strained than the pre-designed $In_{10}As_3V$ molecule. Indeed, the total number of electrons (238) contributing to the top 118 doubly occupied MOs of the pre-designed $Ga_{10}As_3V$ molecule is the same as that contributing to the top 115 doubly occupied MOs of the pre-designed $In_{10}As_3V$ molecule. Given that the Ga-based molecule has 180 less electrons than the In-based one, the above fact signifies that much larger number AOs of deeply lying electrons in Ga atoms of the Ga-based molecule have to be re-configured in response to a disturbance caused by the V atom than that in the case of the In-based molecule.

The V atom in the pre-designed $Ga_{10}As_3V$ molecule accumulates more of re-distributed electron charge of Ga atoms than As atoms in this molecule (Figs. 18c to 18i). In contrast, in the vacuum $Ga_{10}As_3V$ molecule the V atom accumulates less of the electron charge than the As atoms (Figs. 19b to 19f). This is yet another sign that the pre-designed molecule is overly strained, so the V atom has to take more of the Ga electron charge to provide for a stable state (a ROHF triplet; Table II) similar to that of the corresponding vacuum molecule. In the case of much roomier vacuum $Ga_{10}As_3V$ molecule there is less need to accumulate charge near the V atom or to push the charge outside the molecule to stabilize the molecule.

The above analysis leads to a conclusion that dimensions and properties of "holes" mediated by a substitution V-atoms in the zincblende GaAs lattice should be more sensitive to the lattice strain than those in the case of the zibcblende InAs lattice. This phenomenon can be used to develop a sensitive device to measure lattice strain by measuring the hole conductivity, or vice versa.

Several MOs from the HOMO-LUMO regions of the studied $Ga_{10}As_3V$ molecules, both of which are ROHF triplets (Table II), are shown in Figs. 20 and 21. The electronic level structure of the pre-designed molecule retains significant symmetry in the HOMO-LUMO region exhibiting doubly degenerate MOs, and in particular LUMO. Counting from the proper HOMO 121 (which is non-degenerate) toward the core MOs, ELS is 2A (HOMO 120 and MO 119), E (MOs 118 and 117), A (MO 116), E (MOs 115 and 114), E (MOs 113 and 112), A (MO 111), and so on. In the LUMO region, counting from the proper LUMO 121 and up, ELS is E (LUMO 121 and MO 122), E (MOs 123 and 124), A (MO 125), E (MOs 126 and 127), E (MOs 128 and 129), 2A (MO 130 and 131), T (MOs 132, 133 and 134), etc. This is a result of constraining all centers of mass of atoms to their tetrahedral positions in the tetrahedral (pyramidal) symmetry element of the zincblende GaAs lattice. The reduction in symmetry of the electronic charge distribution of this molecule is caused only by replacement of one of Ga atoms by the V atom.

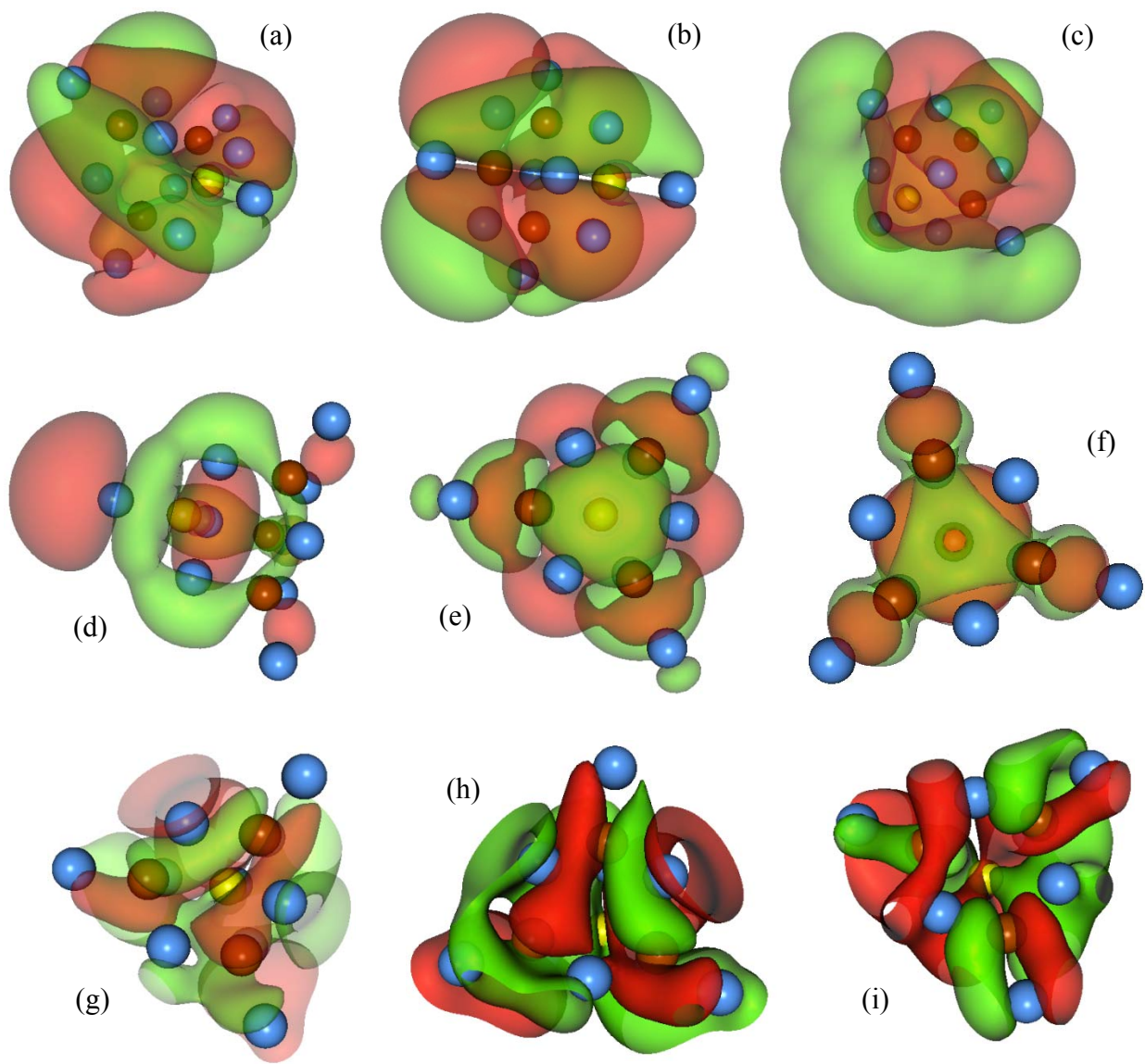

Fig. 20. (Color online) The pre-designed $Ga_{10}As_3V$ molecule. Isosurfaces of the positive (green) and negative (orange) parts of the highest occupied and lowest unoccupied molecular orbits (HOMOs and LUMOs, respectively) corresponding to several isovalues. (a) and (b): HOMOs 117 and 118, isovalue 0.003, respectively; (c) HOMO 119, isovalue 0.001; (d) HOMO 119, isovalue 0.007; (e) and (f): HOMO 120, isovalue 0.007; (g) LUMO 121, isovalue 0.01; (h): LUMO 122, isovalue 0.007, and (i) LUMO 123, isovalue 0.01. Ga atoms are blue, As brown and V yellow. Atomic dimensions are reduced and isosurfaces made transparent to show the structure.

In the pre-designed $Ga_{10}As_3V$ molecule $3d$ AOs of tetra-coordinated vanadium atom always bond it directly to 4 Ga atoms. The arsenic atoms bond the 6 other Ga atoms, and the first 4 Ga atoms bonded to V also bond to 3 Ga atoms from the "arsenic bonding triangle", thus completing an MO (one of the 4 Ga atoms bonded to V is in a pyramid vertex, and does not contribute much to Ga-As π-type bonding). Ga atoms bond both to V and As ones via their $4p$ AOs. In contrast to the case of InAs-based molecules where there were some contributions to bonding from $4d$ AOs of In atoms, there are no contributions to bonding from $3d$ AOs of Ga atoms in the GaAs-based molecules. Arsenic atoms bond through their $4p$ AOs only to Ga atoms, and do not bond to the vanadium atom directly. This arrangement is typical for all MOs in Fig. 20. The (4 + 3) Ga atom $4p$-bonding brings about a strong π-type ligand bonding MOs of this molecule (see Ref. 122 for further discussion of "aromatic" π-type ligand bonding) in the HOMO region. The π-type ligand bonding MOs are responsible for the molecule being a stable ROHF triplet whose OTE (over 1.26 eV) is larger than that of the vacuum $Ga_{10}As_3V$ molecule (about 1.058 eV), and whose minimum of the total energy is almost as deep as that of the vacuum molecule (see Table II).

ELS of the vacuum $Ga_{10}As_3V$ molecule does not exhibit any charge symmetry, being composed only of A-type orbits and showing only a very few spontaneously degenerate MOs of E-type in the higher LUMO region. Several MOs in the near HOMO-LUMO region are depicted in Fig. 21. The type of bonding in this molecule is very similar to that of the pre-designed molecule, with some small deviations. As a rule, 3*d* AOs of its V atom bond 4*p* AOs of the nearest 4 Ga atoms. Four or 3 of these Ga atoms bond another 4 or 3 Ga atoms (Ga-Ga 4*p* – 4*p* bonding), where the 3 Ga are bonded to all 3 As atoms (Ga-As 4*p* – 4*p* bonding). In some cases the Ga atom bonded to V and positioned in the pyramid vertex closest to the V atom may be only weakly bonded to the Ga-As portion of the entire hybrid MO. Such MOs provide for the vacuum molecule being a robust ROHF triplet with a large OTE and a deep energy minimum of its ground state (Table II). The Ga-V bond length in this molecule takes only 2 values: 2.505 Å and 2.998 Å. The bond length of the ligand Ga-Ga bonding also takes two values: 4.046 Å and 4.532 Å, while Ga-As bonding is more flexible, so the Ga-As bond may be 2.487 Å, 2.519 Å and 2.835 Å. Similar to the case of the pre-designed molecule, in the vacuum molecule vanadium and arsenic atoms do not bond directly.

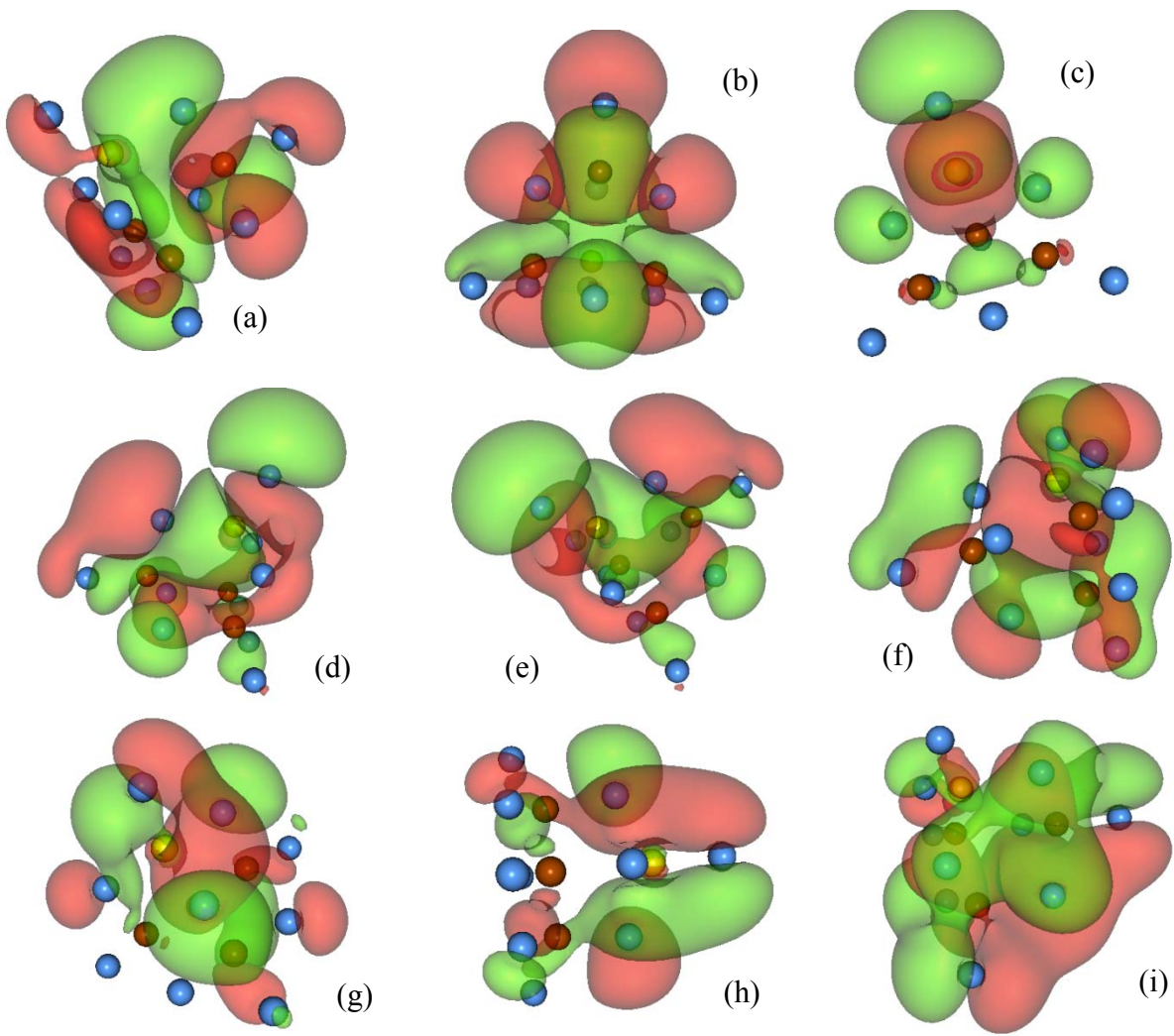

Fig. 21. (Color online) The vacuum $Ga_{10}As_3V$ molecule. Isosurfaces of the positive (green) and negative (orange) parts of the highest occupied and lowest unoccupied molecular orbits (HOMOs and LUMOs, respectively) corresponding to several isovalues. (a) and (b): HOMO 118, isovalues 0.01 and 0.005, respectively; (c) HOMO 119, isovalue 0.01; (d) and (e): HOMO 120, isovalue 0.005; (f): LUMO 121, isovalue 0.015; (g) and (h): LUMO 122, isovalues 0.01 and 0.01, respectively, and (i) LUMO 123, isovalue 0.01. Ga atoms are blue, As brown and V yellow. Atomic dimensions are reduced and isosurfaces made transparent to show the structure.

Compared to InAs-based molecules with one V atom that are pre-designed ROHF nonet and vacuum ROHF pentet, GaAs-based ones are more stable, being ROHF triplets. This phenomenon is due to extraordinary stability of bonding MOs of this molecule that possess a large π-type Ga-ligand (122) and Ga-As bonding contributions.

Being ROHF triplets, both $Ga_{10}As_3V$ molecules have "ferromagnetic" arrangement of aligned uncompensated spins (Fig. 22) contributed by electrons in 3*d* AOs of Ga atoms, with a very small contribution of V atoms and even smaller contributions of As atoms. The SDD values of these molecules are about an order of magnitude smaller than those of the InAs-based molecules with one V atom discussed in Sec. 4, that are higher ROHF spin multiplets. Correspondingly, the magnetic moment of the GaAs-based molecules with one V atom is significantly smaller (3 $\mu_B$) than the smallest of the magnetic moments of InAs-based molecules 5 $\mu_B$) with one V atom. Thus, for DMS applications with an emphasis on the large magnetic moment $In_{10}As_3V$ molecules may be a better choice.

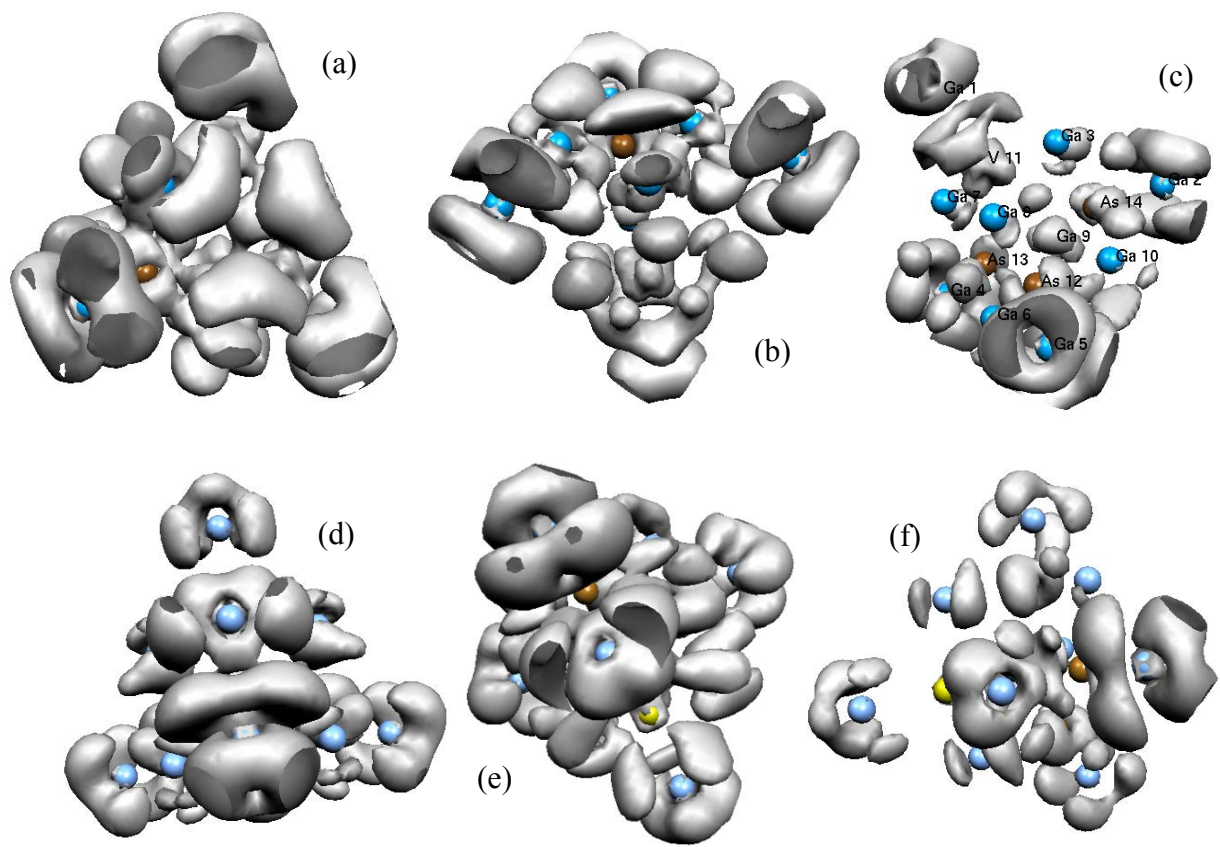

Fig. 22. Isosurfaces of the spin density distribution (SDD) of the pre-designed [(a) to (c)] and vacuum [(d) to (f)] $Ga_{10}As_3V$ molecules corresponding to the fractions (a) 0.0005, (b) 0.0007, (c) 0.001, and (d) 0.0007, (e) 0.0005, (f) 0.001 of the respective SDD maximum values (not shown). Indium atoms are yellow, As red and Mn blue. All atomic dimensions are reduced to show the SDD surface structure.

# 6. InAs - AND GaAs - BASED MOLECULES WITH TWO VANADIUM ATOMS.

Pre-designed and vacuum $In_{10}As_2V_2$ and $Ga_{10}As_2V_2$ molecules have been virtually synthesized using the same procedures utilized for virtual synthesis of $In_{10}As_2V$ and $Ga_{10}As_2V$ molecules (see Sections 1, 4 and 5 of this Chapter), with the only difference that this time two As atoms have been replaced by two V atoms in each of the pre-designed $In_{10}As_4$ and $Ga_{10}As_4$ tetrahedral pyramids (see Chapter 4) of the zincblende InAs and GaAs lattices. After the replacement, the $In_{10}As_2V_2$ and $Ga_{10}As_2V_2$ pyramidal structures were optimized while the centers of mass of their atoms were constrained to the same positions in space. The result of this optimization is two pre-designed molecules $In_{10}As_2V_2$ and $Ga_{10}As_2V_2$. Then the initial structures were optimized again, this time when the constraints applied to the atomic positions were lifted, producing two so-called vacuum molecules $In_{10}As_2V_2$ and $Ga_{10}As_2V_2$. The structure of all four molecules is pictured in Fig. 23. Because of the spatial constraints applied to the atomic positions, both of the pre-designed molecules (Figs. 23a and 23d) retain their pyramidal shape. When the constrains are lifted, only the vacuum $Ga_{10}As_2V_2$ molecule (Figs, 23e and 23f) appears pyramidal, while the vacuum $In_{10}As_2V_2$ molecules loses any resemblance to the tetrahedral pyramid of its parent structure. These phenomena, of course, follow from the relative sizes of the participating atoms and the number of electrons they possess. Thus, the V atom is smaller and has fewer electrons than As and Ga atoms, so after the replacement of 2 As atoms by 2 V atoms in the original $Ga_{10}As_4$ pyramid (see Chapter 3 for details on this structure) the pre-designed molecule $Ga_{10}As_2V_2$ is somewhat loose. When spatial constraints on atomic positions are lifted, the atoms move from their original positions, but only by several tenths of Angstrom. Thus, visually, the vacuum $Ga_{10}As_2V_2$ molecule appears pyramidal. In the case of $In_{10}As_2V_2$, the V atom is much smaller and has much fewer electrons that the In atom. After the replacement of two As atoms by two V ones the original $In_{10}As_4$ structure becomes too loose, as the absence of 20 electrons that occur after the replacement creates a large charge and mass distribution imbalance. Thus, when the constraints applied to atomic positions are lifted, atoms in the original structure adjust dramatically in response to the charge imbalance. As a result, the vacuum $In_{10}As_2V_2$ molecule assumes entirely a shape entirely different from that of its parent pyramidal structure and becomes a stable ROHF singlet. In contrast, its pre-designed counterpart has spin multiplicity 11, and thus is an unstable molecule (see Table II) in the framework of ROHF approximation used here. The CDDs and MEPs of the studied molecules with two V atoms reflect characteristic features of their composition and shape (Figs. 24 and 25).

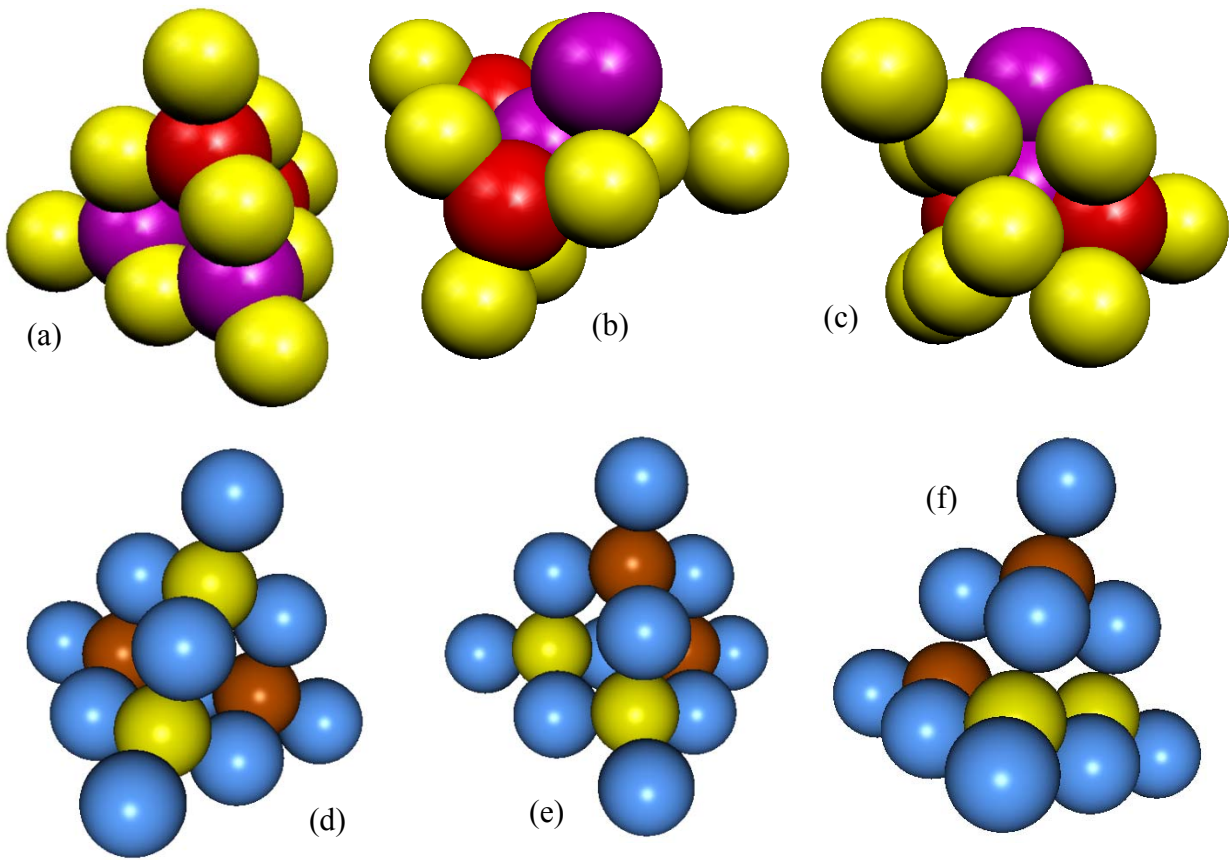

Fig. 23.  (Color online) The pre-designed (a) and vacuum [(b) and (c)] $In_{10}As_2V_2$ molecules, and the pre-designed (d) and vacuum [(e) and (f)] $Ga_{10}As_2V_2$ molecules, respectively. Atomic dimensions approximately correspond to the atoms' covalent radii. In the $In_{10}As_2V_2$ molecules In atoms are yellow, As red and V blue. In the $Ga_{10}As_2V_2$ molecules Ga atoms are blue, As brown, and V yellow.

Comparison of MEP surfaces of the pre-designed and vacuum $In_{10}As_2V_2$ molecules in Fig. 24 obtained for similar CDD isovalues reveals strikingly different electrostatics of these molecules. Thus, the MEP surface values of the pre-designed molecule for the CDD isovalue 0.001 ran from highly negative to positive ones (Fig. 24a), while for the close CDD isovalue 0.0007 the MEP values of the vacuum molecule are uniformly close to 0 (Fig. 24d). Thus, in the case of the pre-designed molecule, at large distances from the molecular "surface" there still exists large MEP fluctuations that reflect instability of this molecule. In contrast, the vacuum molecule is stable exhibiting text-book MEP values for large separations from the molecular "surface". Similar conclusions follow from comparison of MEP values in Figs. 24b and 24e for the same CDD isovalue 0.01, and those in Figs. 24c and 24f for close CDD isovalues 0.08 and 0.1, respectively. Comparison of the ground state energy values of these molecules reveals that the ground state energy values differ by about 1 H (Table II). This difference lies at the limit of the best total energy evaluation accuracy of the RHF-ROHF approximation method. The dipole moment of the pre-designed molecule is also somewhat larger than that of the vacuum one (Table II). These facts indicate that the pre-designed molecule that realizes a local minimum of the total energy of the $In_{10}As_2V_2$ atomic cluster (corresponding to the tetrahedral symmetry spatial constraints applied to the cluster's atoms) is not stable and may not be realizable experimentally.

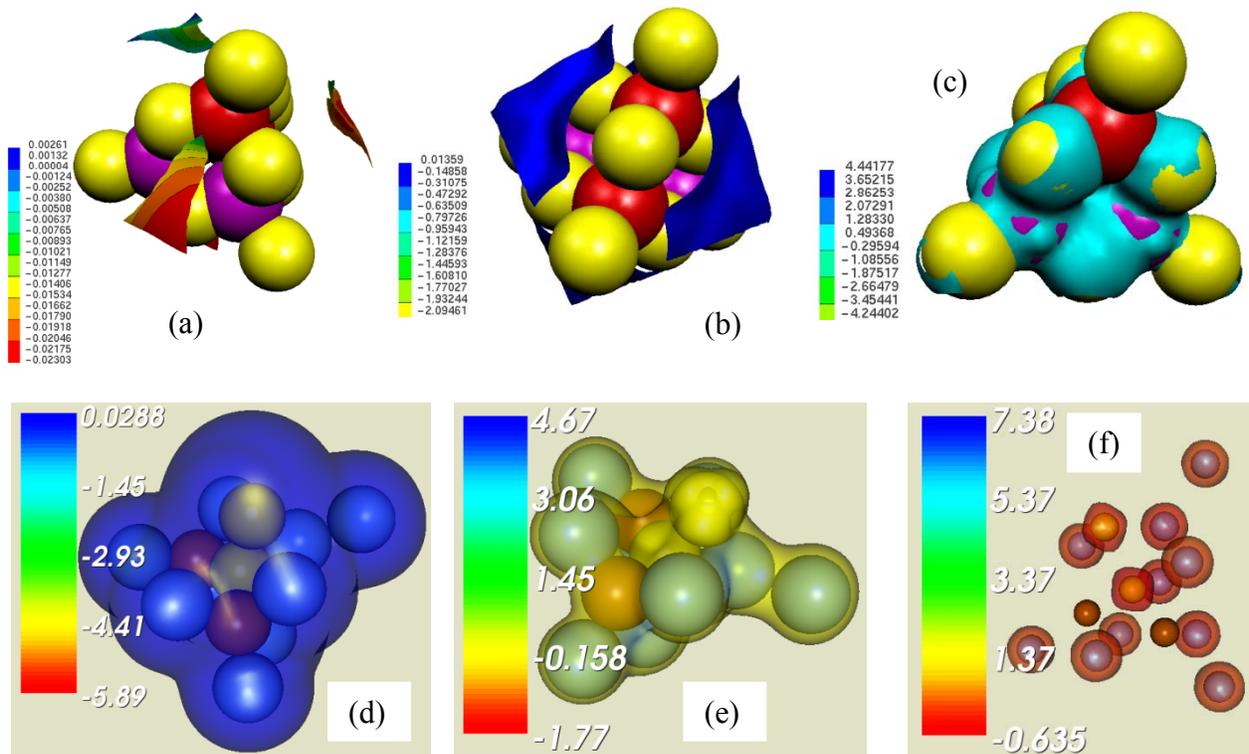

Fig. 24. (Color online) The molecular electrostatic potential (MEP) of the pre-designed [(a) to (c)] and vacuum [(d) to (f)] $In_{10}As_3V_2$ molecules for several isosurfaces of the corresponding CDDs calculated for the following fractions (isovalues) of the CDD maximum values 3.44417 and 3.01378 (in arbitrary units), respectively: (a) 0.001, (b) 0.01, (c) 0.08, (d) 0.0007, (e) 0.01, and (f) 0.1. The color coding scheme for MEP surfaces is shown in each figure. In the pre-designed molecule [(a) to (c)] In atoms are yellow, As red and V purple. In the vacuum molecule [(d) to (f)] In atoms are blue, As brown and V yellow. In (a) to (e) atomic dimensions of In atoms are somewhat smaller than those defined by the In atom's covalent radius. In (a) to (c) the dimensions of V and As atoms are enlarged. In (f) all atomic dimensions are significantly reduced to show the MEP surface structure. In (d) to (f) MEP surfaces are semi-transparent to reveal the structure.

Several MEP surfaces of $Ga_{10}As_2V_2$ molecules for several CDD isovalues are shown in Fig. 25. In contrast to those of the $In_{10}As_2V_2$ molecules, the MEP surfaces of $Ga_{10}As_2V_2$ ones are very similar. This is a consequence of the fact that both $Ga_{10}As_2V_2$ molecules are close ROHF spin multiplets (see Table II). On the other hand, both the pre-designed ROHF $Ga_{10}As_2V_2$ septet and vacuum $Ga_{10}As_2V_2$ nonet are higher excited states, and are not typical for the ground state of the majority of small stable molecules. Moreover, the pre-designed septet is more stable than the vacuum nonet, indicating that the calculated minimum of the total energy in the case of the vacuum nonet may not be a global minimum. Indeed, comparing for example, MEP surfaces of these molecules depicted in Figs. 25 c and 25f, one can see that the pre-designed molecule has larger areas of delocalized electron charge shared between Ga, V and As atoms than the pre-designed one. Most likely, the vacuum nonet is a realization of a local minimum of energy of $Ga_{10}As_2V_2$ atomic cluster. This consideration is further supported by comparison of the ground state energies of these molecules (Table II) that are lying within the error brackets of the RHF-ROHF approximation. Indeed, the difference in the ground state energy values of these molecules is only 0.055 H, while the best possible total energy evaluation accuracy of the used calculation method is a few Hartrees. The dipole moment of the vacuum molecule is considerably smaller than that of the pre-designed one (Table II). This further supports the conclusion that the vacuum nonet is a local minimum of the total energy of the unconstrained $Ga_{10}As_2V_2$ cluster that correspond to a more uniform electron charge distribution than that specific to the local minimum of the constrained pre-designed molecule.

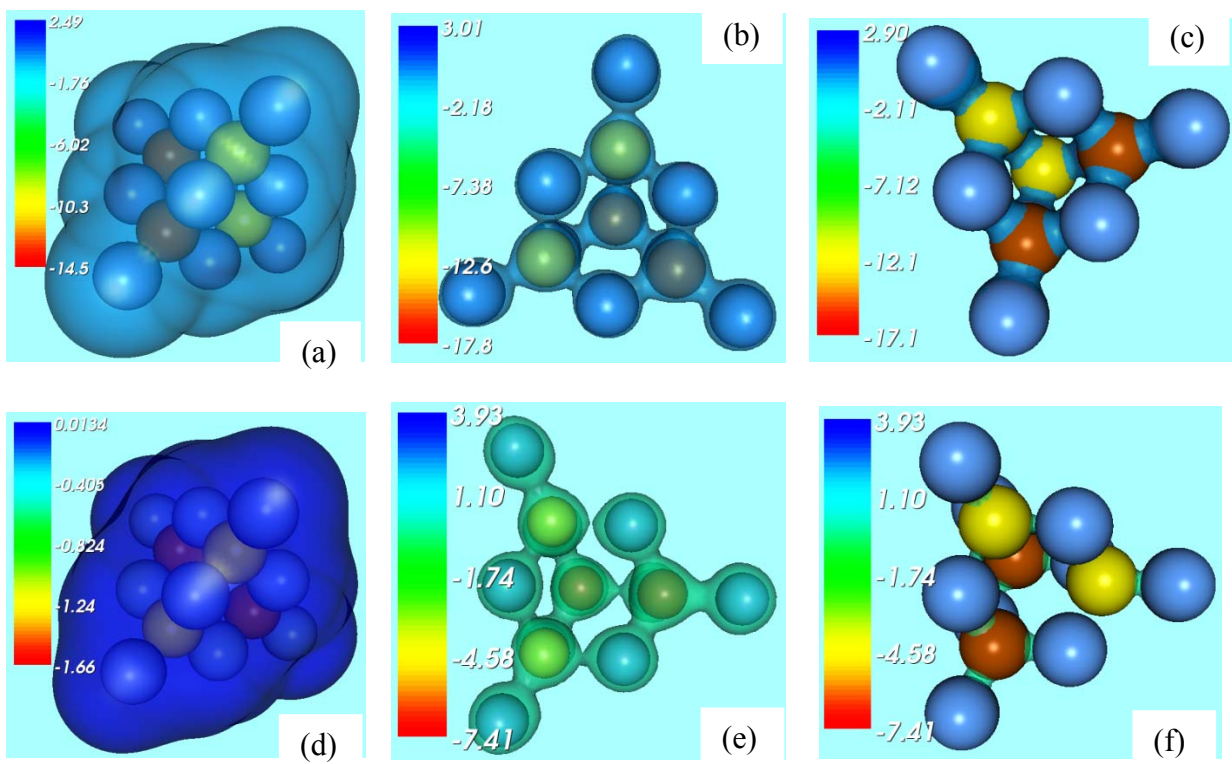

Fig. 25. (Color online) The molecular electrostatic potential (MEP) of the pre-designed [(a) to (c)] and vacuum [(d) to (f)] $Ga_{10}As_3V_2$ molecules for several isosurfaces of the corresponding CDDs calculated for the following fractions (isovalues) of the CDD maximum values 9.46923 and 10.37640 (in arbitrary units), respectively: (a) 0.007, (b) 0.01, (c) 0.08, (d) 0.005, (e) and (f): 0.1. The color coding scheme for MEP surfaces is shown in each figure. Ga atoms are blue, As brown and V yellow. In (a), (c), (d) and (f) atomic dimensions roughly correspond to those defined by the atoms' covalent radii. In (b) to (e) all atomic dimensions are reduced to show the MEP surface structure. In all cases but (c) MEP surfaces are semi-transparent to reveal the structure.

Molecular orbitals of the molecules with two V atoms are shown in Figs. 26 and 27. In support of the conclusion regarding instability of the pre-designed $In_{10}As_2V_2$ molecule discussed above, MOs of this molecule in the near HOMO-LUMO region (Figs. 26a and 26b) feature antibonding parts contributed to by $3d$ AOs of In and V atoms, and bonding $sp$ - type portions due to As-In bonding, and π-type portions of In ligand bonding. The LUMO (Fig. 26c) of this molecule is a bonding orbital realized via hybrid type $pd$-type In-V, $p$-type In-As, and π-type In-In bonding. MOs of the vacuum $In_{10}As_2V_2$ molecule in its HOMO-LUMO region (Figs. 26d to 26f) are bonding hybrid MOs with significant portions of $pd$ In-V bonding, and As-mediated π-type In ligand bonding.

In the case of $Ga_{10}As_2V_2$ molecules (Fig. 27), all MOs in the HOMO region and the LUMOs are bonding hybrid MOs contributed to by (1) $pd$-type Ga-V bonding due to $3d$ AOs of V and $4p$ AOs of Ga, (2) $4p$-type As-Ga bonding, and (3) π-type Ga ligand bonding.

Magnetic properties of the studied InAs- and GaAs-based molecules are illustrated in Fig. 28. An important finding is that replacement of an As atom by the second vanadium one in the studied $In_{10}As_3V$ and $Ga_{10}As_3V$ molecules does not ensure an increase in the magnetic moment of the obtained molecules, because it generally destabilizes the original molecules. At this point in studies, it seems that the pre-designed $Ga_{10}As_2V_2$ septet (Figs. 28c and 28d) may be experimentally realized in quantum confinement. Indeed, this molecule has a deep ground state energy minimum (Table II) that is about 720 H deeper than that of the vacuum $In_{10}As_3Mn$ molecule, which is the only other relatively stable molecule with a large uncompensated magnetic moment $7\mu_B$. At the same time, the vacuum $In_{10}As_3Mn$ molecule has been computationally synthesized in the absence of any spatial constraints applied to its atoms, and such conditions are not exactly realizable in experiment. Moreover, the pre-designed $In_{10}As_3Mn$ molecule is "antiferromagnetic" with zero total uncompensated magnetic moment. Thus, it seems likely that application of spatial constraints to positions of atoms in $In_{10}As_3Mn$ cluster atoms in the process of experimental synthesis of $In_{10}As_3Mn$ molecules (let alone films) may lead to a molecule with a smaller uncompensated magnetic moment than that of the vacuum $In_{10}As_3Mn$ one analyzed above. Thus, among InAs- and GaAs-based molecules with two V substitution atoms only the pre-designed $Ga_{10}As_2V_2$ may be of interest for DMS applications. Indeed, as already noted, the pre-designed $In_{10}As_2V_2$ molecule is a ROHF singlet, while the vacuum $In_{10}As_2V_2$ (Figs. 28a and 28b) and $Ga_{10}As_2V_2$ (Figs. 28e and 28f) molecules may not be realizable experimentally.

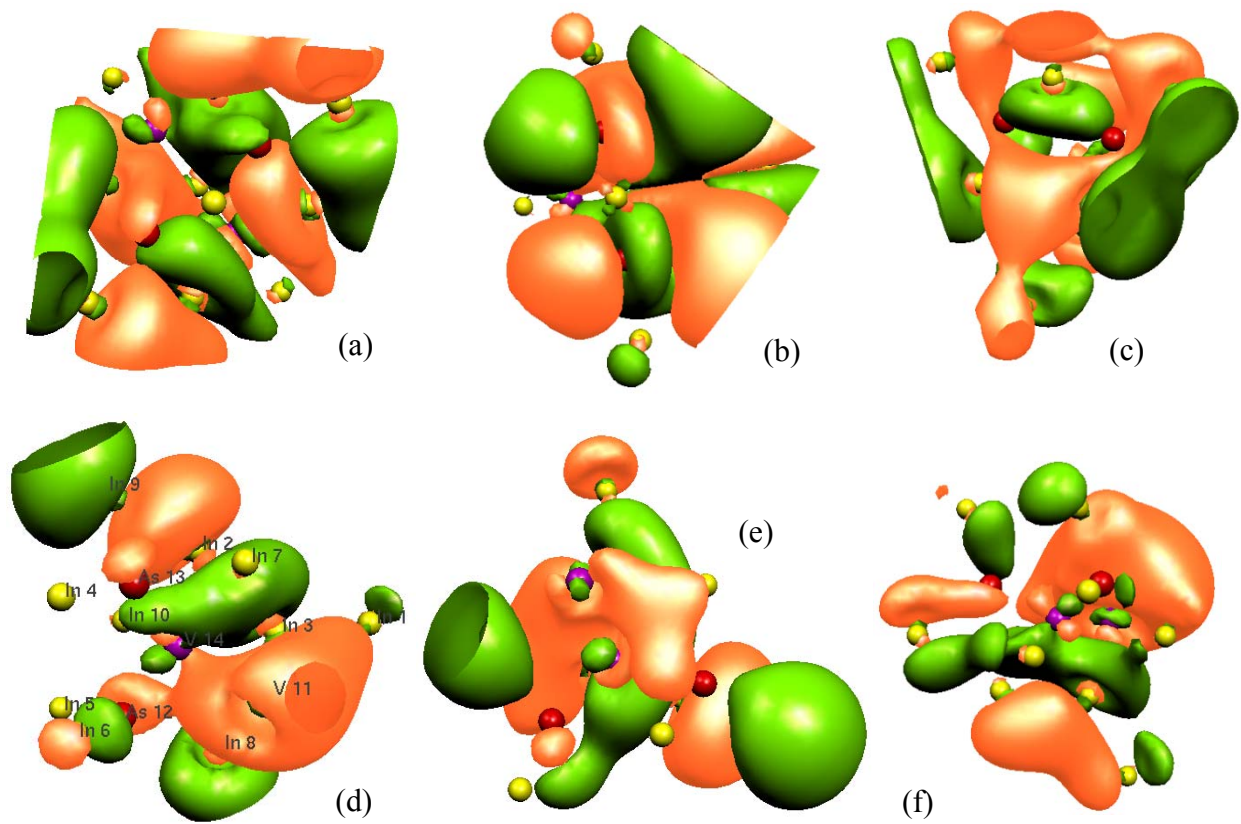

Fig. 26. (Color online) Isosurfaces of the positive (green) and negative (orange) parts of the highest occupied and lowest unoccupied molecular orbits (HOMOs and LUMOs, respectively) corresponding to several isovalues. The pre-designed $In_{10}As_2V_2$ molecule: (a) HOMO 127, isovalue 0.01; (b) HOMO 128, isovalue 0.001; (c) LUMO 129, isovalue 0.015. The vacuum $In_{10}As_2V_2$ molecule: (d) HOMO 122, isovalue 0.015; (e): HOMO 123, isovalue 0.01; (f): LUMO 124, isovalue 0.015. Indium atoms are yellow, As red and V purple. Atomic dimensions are reduced to reveal the isosurface structure.

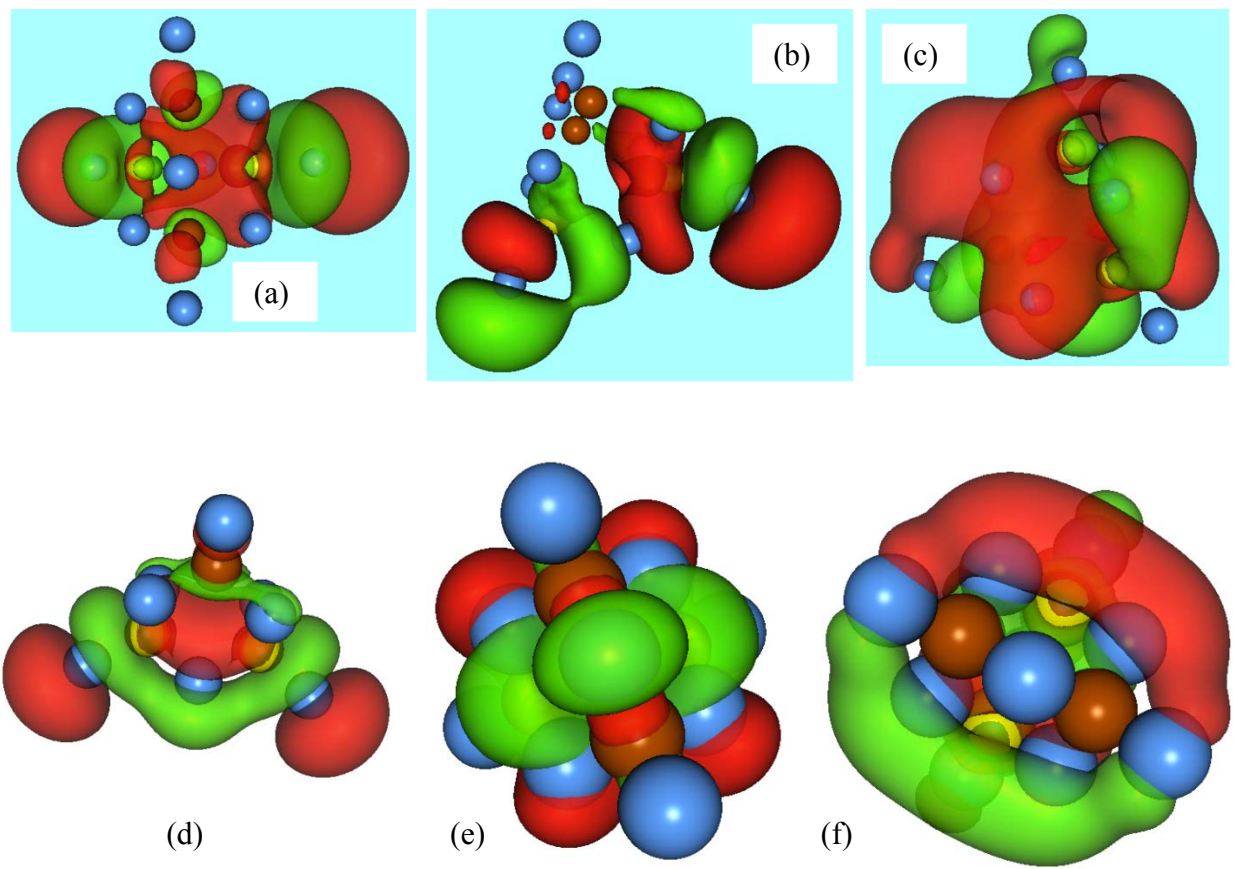

Fig. 27. (Color online) Isosurfaces of the positive (green) and negative (orange) parts of the highest occupied and lowest unoccupied molecular orbits (HOMOs and LUMOs, respectively) corresponding to several isovalues. The pre-designed $Ga_{10}As_2V_2$ molecule: (a) HOMO 125, isovalue 0.008; (b) HOMO 126, isovalue 0.01; (c) LUMO 127, isovalue 0.008. The vacuum $Ga_{10}As_2V_2$ molecule: (d) HOMO 126, isovalue 0.01; (e): HOMO 127, isovalue 0.01; (f): LUMO 128, isovalue 0.01. Ga atoms are blue, As brown and V yellow. In (a) to (d) Atomic dimensions are reduced to reveal the isosurface structure. In (e) and (f) the atomic dimensions are reduced only slightly. Isosurfaces are semi-transparent to reveal the structure.

## 7. CONCLUSIONS

In this Chapter first-principle, quantum statistical mechanical methods, RHF and ROHF, were used to synthesize virtually 10 molecules optimized from In, Ga and As atoms in arrangements that reflect tetrahedral symmetries of the zincblende InAs and GaAs bulk lattices, or are derivable from such arrangements by the total energy minimization. The above discussion follows initial steps toward realization of a program of virtual studies of semiconductor atomic

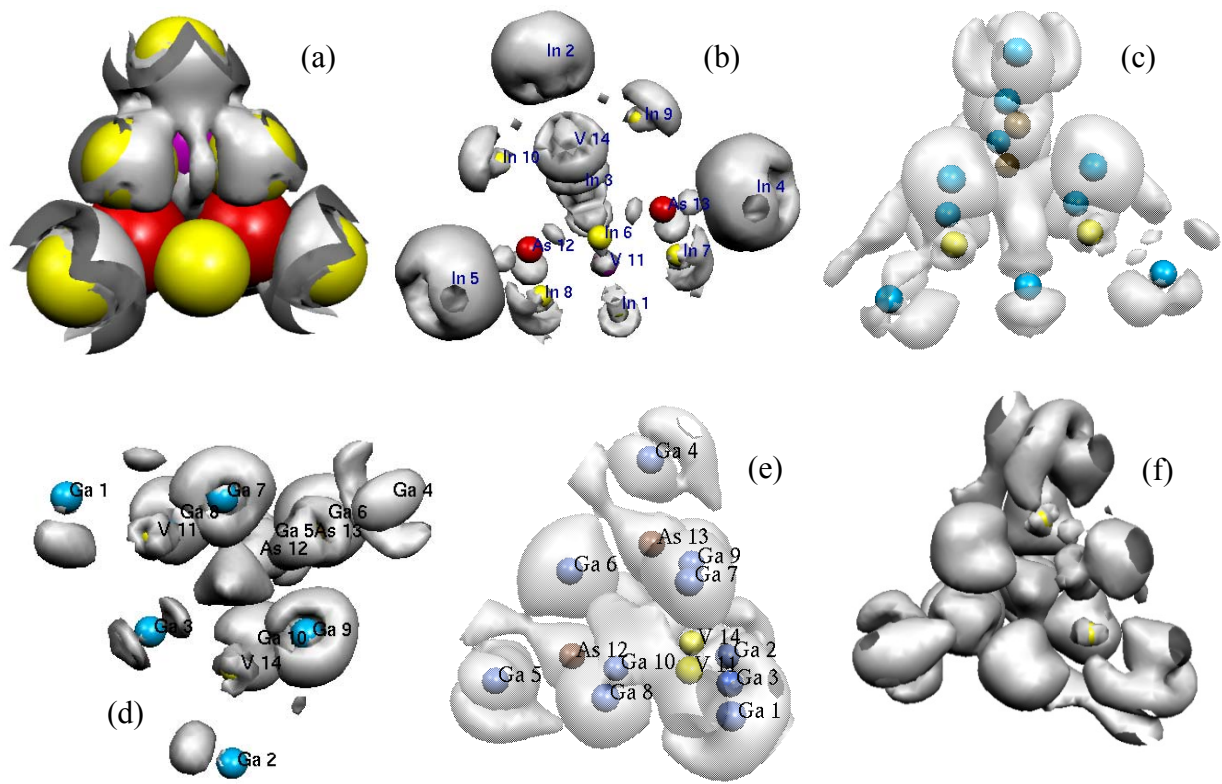

Fig. 28. Isosurfaces of the spin density distributions (SDDs) of the pre-designed $In_{10}As_2V_2$ [(a) and (b)] and $Ga_{10}As_2V_2$ [(c) and (d)] molecules, and the vacuum molecule $Ga_{10}As_2V_2$ [(e) and (f)], corresponding to the fractions (isovalues): (a) 0.001, (b) 0.004, (c) 0.002, (d) 0.003, (e) 0.002 and (f) 0.002 of the respective SDD maximum values (not shown). In (a) and (b) indium atoms are yellow, As red and Mn blue. In (c) to (f) Ga atoms are blue, As brown and V yellow. In (b) to (f) all atomic dimensions are reduced to show the SDD surface structure. In (a) atomic dimensions are roughly correspond to those defined by the atoms' covalent radii.

clusters and nanoscale systems introduced in Refs. 119 – 121, 123, 124 and related publications.

It is necessary to underline here that much more research is needed before any final quantitative conclusions concerning the systems discussed here are reached. It is well known (see Chapter 3) that RHF and ROHF methods overestimate OTE values and do not provide accurate molecular orbitals for atomic clusters so optimized. However, it is also known that these methods permit to obtain important information and data to enable further theoretical studies and to guide experimental developments. In particular, CI, MCSCF, MP-2 and other subsequent and progressively more accurate, first principle approximation methods use RHF or ROHF MOs as input, and the RHF/ROHF ground state energy, CDD and SDD data provide important information and ideas for prospective applications.

Several questions important for DMS applications concern the nature of magnetism and band structure the band structure exhibited by DMS, and the location of the holes mediated by impurity atoms (see Sec. 1 for references). Analysis of RHF/ROHF data discussed in this Chapter provides some important insights that may answer the above questions posed by experimentalists studying DMS systems.

1. In the case of nanoscale InAs- and GaAs-based DMS systems (such as thin films, small QD and QWs, etc.) with at least one linear dimension in the range of a few nanometers the ELS structure is defined by quantum confinement and Coulomb interaction effects, and cannot be described in terms of impurity-driven modification of the band systems of the parent InAs and GaAs bulk lattices. Moreover, such ELSs may only remotely remind a band structure, especially in the case of small QDs. Rather, ELS in such cases is similar to that of molecules and is formed by all participating atoms. Perturbation theory-based approaches no not work for such strongly correlated electron systems.

2. Magnetism in DMS systems is derived from electron charge and spin re-distribution in a broad vicinity of impurity atoms that may include over 10 lattice atoms surrounding the impurity atoms (and not necessarily centered at the impurity atoms). This electron charge and spin re-distribution is a response of the nearby lattice atoms to the impurity-based disturbance of the electron charge of the lattice atoms. In the majority of the studied cases, the non-zero total spin magnetic moment arises from uncompensated electron spins of $4d$ In or Ga electrons in the process of such electron charge re-distribution.

3. In the studied cases, the process of the electron charge re-distribution leads to charge delocalization about up to 13 atoms neighboring the substitution impurity atom(s). The re-distributed charge sets in highly hybridized, bonding molecular orbits that possess three major types of contributions. In the majority of the studied cases, one of two major contributions come from $pd$-type hybrid bonding between of Mn or V atoms, and 3 to 4 ligand atoms (In or Ga): Mn or V contribute through their $3d$ AOs, and In or Ga atoms through their hybridized $4p$ AOs that develop in response to the presence of $3d$ AOs of Mn or V atoms. The other major contribution to bonding MOs comes from hybrid $p$-bonding of Ga or In with As atoms. This later contribution also mediates ligand bonding between up to 4 In or Ga atoms. These results confirm and enrich the major idea of the $p$-$d$ Zener model. Also, the obtained results do not reveal any significant $4p$-bonding between Mn and As atoms in InAs-based molecules, in contrast to suggestion of Ref. 79. Instead, there are some contributions to bonding MOs from $5p$-$4d$ bonding between different In atoms in such molecules ($pd$ ligand bonding).

4. The total uncompensated magnetic moment is delocalized over the entire molecules, but is at maximum in the vicinity of In or Ga atoms. Because RHF/ROHF MOs may not be quite accurate, especially in the case where many-electron atoms, such as In, Ga and As, are involved, further detailed studies of the considered systems by CI, MCSCF and MP-2 methods are necessary to identify quantitatively mechanisms of "generation" of the uncompensated magnetic moment in the studied molecules. At this stage, it is clear that uncompensated total magnetic moment appears in response to charge disturbance introduced by $3d$ electrons of Mn or V atoms. However, only in two of the studied cases the uncompensated magnetic moment is spread about the geometrical centers of the molecular structures (but not in the vicinity of Mn or V atoms). In the other 8 studied cases the total magnetic moment is delocalized further including all In or Ga atoms. It is also clear that quantitatively, the total magnetic moment is not equal to what could have been contributed by $3d$ electrons of V or Mn atoms alone, as their contributions are small. The spin multiplicity of the virtually synthesized molecules range from 1 to 11, which signifies that many more electron spins residing on Ga or In atoms are involved. Indeed, calculations of each of the topmost occupied RHF/ROHF MOs for the studied molecules involves about 120 electrons of the parent atoms, instead of involving only "valence" electrons of the atoms.

5. The electron charge delocalization about many atoms surrounding an impurity atom creates regions of electron charge deficit with the major portion surrounding the "surfaces" of the studied molecule on outside of and inside the molecule structures. Although the total charge of the molecules is zero, such "shells" of the electron charge deficit behave as delocalized and polarized positive charge when external electromagnetic fields ate applied. They separate the electron charge localized deeper inside the molecular structure from that pushed outside the molecular "surface". The electron charge deficit "shells" are physical realization of idealized, positively charged "holes" of the semi-phenomenological band theory of semiconductors. Such "shells" – holes – embrace from about 8 to 14 atoms in the studied cases, and are larger in size for In-based molecules, exceeding 1 nm in linear dimensions. Due to the total non-zero magnetic moment of such molecules, the holes are spin-polarized, as they are regions of deficit if the spin-polarized electron charge.

6. Considering that linear dimensions of the holes mediated by substitution impurity atoms in the studied cases exceed 1 nm, in thin DMS films and other nanosystems with at least one linear dimension in the range of a few nanometers such holes may include atoms of other systems spatially confining a DMS nanosystem. Moreover, if a suitably directed electromagnetic field gradient is applied, these "shells" of electron charge deficit will move in response to the electron motion, and may leave the DMS nanosystem of their origin.

7. With a change in thermodynamic conditions, such as temperature or pressure applied to a DMS nanosystem, structural changes may provoke another electron charge re-distribution that will affect the shape of the holes, and thus affest the impurity-mediated magnetic moment. Among the studied molecules only two ($Ga_{10}As_3V$) ones did not change their spin multiplicity in response to a change in conditions of their synthesis. The rest of the studied molecules changed values of their total uncompensated magnetic moments in response to a change in synthesis conditions, and two molecules ($In_{10}As_3Mn$ and $In_{10}As_2V_2$) became "antiferromagnetic" RHF singlets with the total magnetic moment equal to zero. These results identify a mechanism behind transitions from a ferromagnetic state to a non-magnetic state in some DMS thin film systems observed in recent experiments (see Sec. 1 for references and a further discussion of these experimental observations).

8. All calculations of this chapter are done for systems at zero absolute temperature. However, the OTE of all virtually synthesized, stable molecules (including even septets, Table II) are much larger than any temperature-derived contributions at room temperature. Thus, the obtained results are likely to be relevant, in some cases even quantitatively, to the studied systems at room temperature.